\documentclass{pasj00}
\usepackage[varg]{txfonts}

\SetRunningHead{K. Sato et al.}{
Suzaku observations of the Sculptor Supercluster}

\title{
Study of the Intracluster and Intergalactic Medium \\
in the Sculptor Supercluster with Suzaku
}
\author{
 Kosuke \textsc{Sato},\altaffilmark{1}
 Richard \textsc{L.~Kelley},\altaffilmark{2}
 Yoh \textsc{Takei},\altaffilmark{3}
 Takayuki \textsc{Tamura},\altaffilmark{3}\\
 Noriko \textsc{Y.~Yamasaki},\altaffilmark{3}
 Takaya \textsc{Ohashi},\altaffilmark{4}
 Anjali \textsc{Gupta},\altaffilmark{5} 
and  Massimiliano \textsc{Galeazzi}\altaffilmark{5} 
}
\altaffiltext{1}{
Kavli Institute for Astrophysics and Space Research,
Massachusetts Institute of Technology, \\
77 Massachusetts Ave., Cambridge, MA 02139, USA}
\email{k\_sato@mit.edu}
\altaffiltext{2}{
NASA/Goddard Space Flight Center, Greenbelt, MD 20771, USA}
\altaffiltext{3}{
Institute of Space and Astronautical Science (ISAS),
Japan Aerospace Exploration Agency, \\
3-1-1 Yoshinodai, Chuo-ku, Sagamihara, Kanagawa 252-5210}
\altaffiltext{4}{
Department of Physics, Tokyo Metropolitan University,
 1-1 Minami-Osawa, Hachioji, Tokyo 192-0397}
\altaffiltext{5}{
Physics Department, University of Miami, Coral Gables, FL 33124, USA}
\Received{2010 July 14}
\Accepted{2010 September 15}
\Published{---}

\begin{document}
\maketitle

\begin{abstract}
  We studied the high temperature plasma in the direction of the
  Sculptor supercluster at $z=0.108$ with Suzaku.  Suzaku carried out
  four observations in the supercluster: namely, A2811, A2811 offset,
  A2804, A2801 regions in 2005 Nov.--Dec., including the regions
  beyond the virial radii of these clusters.  The study needed precise
  background estimation because the measured intensity of the
  redshifted lines, especially those from oxygen, were strongly
  affected by the the Galactic emission.  The spectra taken in the
  regions outside of the virial radii of the member clusters were used
  as the background which included both the Galactic and Cosmic X-ray
  Background (CXB) components.  We also used the background data which
  were taken near the Sculptor supercluster.  Temperature and metal
  abundance profiles were determined to the virial radii of the member
  clusters, and then we searched for the oxygen line emission in the
  region outside of the virial radii of the clusters.  As a result,
  the temperature of the clusters decreased toward the virial radii,
  and the spectral fits for the filament region did not require extra
  component other than the Galactic and CXB components.  We
  constrained the intensities of O\emissiontype{VII} and
  O\emissiontype{VIII} lines to be less than 8.1 and 5.1 $\times10^{-8}$ photons
  cm$^{-2}$ s$^{-1}$ arcmin$^{-2}$, respectively, as 2$\sigma$ upper
  limits.  The intensity of O\emissiontype{VII} indicates $n_{\rm
    H}< 1.6\times 10^{-5}$ cm$^{-3}$ ($Z/0.1~Z_{\odot}$)$^{-1/2}$
  ($L/25~{\rm Mpc}$)$^{-1/2}$, which corresponds to an over density,
  $\delta<60$ ($Z/0.1~Z_{\odot}$)$^{-1/2}$ ($L/25~{\rm
    Mpc}$)$^{-1/2}$\@.
\end{abstract}

\section{Introduction}
\label{sec:intro}
\begin{table*}
\caption{Basic properties of the clusters.}
\label{tab:1}
\begin{center}
\begin{tabular}{lcccc} \hline 
 & redshift & virial radius$^{\ast}$ & Average temperature & $N_{\rm H}$ \\
&& (Mpc/arcmin) & $k\langle T\rangle$ (keV) & (cm$^{-2}$) \\
\hline 
 A2811 & 0.1079 & 1.96/\timeform{16'5}& 5.0 & 1.54$\times10^{20}$\\
 A2804 & 0.108 & 0.88/\timeform{7'3}& 1.0 & 1.78$\times10^{20}$\\
 A2801 & 0.108 & 1.25/\timeform{10'5}& 2.0 & 1.70$\times10^{20}$\\
\hline\\[-1ex]
\multicolumn{5}{l}{\parbox{0.6\textwidth}{\footnotesize 
\footnotemark[$\ast$]
At a redshift of $z=0.108$, $1'$ corresponds to 119~kpc,
and the virial radii, $r_{\rm 180} = 1.95\;
h_{100}^{-1}\sqrt{k\langle T\rangle/10~{\rm keV}}$~Mpc
\citep{markevitch98}}}\\
\end{tabular}
\end{center}
\end{table*}
\begin{table*}[t]
\caption{Suzaku Observation logs of the Sculptor supercluster.}
\label{tab:2}
\begin{tabular}{lccccc} \hline 
Region & Seq. No. & Obs. date & \multicolumn{1}{c}{(RA, Dec)$^\ast$} &Exp.&After screening \\
&&&J2000& ksec &(BI/FI) ksec \\
\hline 
 A2811 & 800005010 & 2005-11-28T04:01:40 & (\timeform{00h41m52.9s},
 \timeform{-28D33'19''})& 30.6& 29.3/29.2\\
 A2811 offset & 800006010 & 2005-11-28T21:10:16 & (\timeform{00h40m42.7s},
 \timeform{-28D44'00''})& 33.9& 32.4/32.3\\
 A2804 & 800007010 & 2005-12-02T08:51:05 & (\timeform{00h39m35.2s},
 \timeform{-28D54'35''})& 31.6& 30.0/29.8\\
 A2801 & 800008010 & 2005-11-27T08:30:47 & (\timeform{00h38m31.3s},
 \timeform{-29D02'48''})& 34.8& 33.4/29.5\\
\hline\\[-1ex]
\multicolumn{6}{l}{\parbox{0.9\textwidth}{\footnotesize 
\footnotemark[$\ast$]
Average pointing direction of the XIS, written in the 
RA\_NOM and DEC\_NOM keywords of the event FITS files.}}\\
\end{tabular}
\end{table*}
\begin{figure*}[!ht]
\begin{center}
\begin{minipage}{0.45\textwidth}
\FigureFile(\textwidth,\textwidth){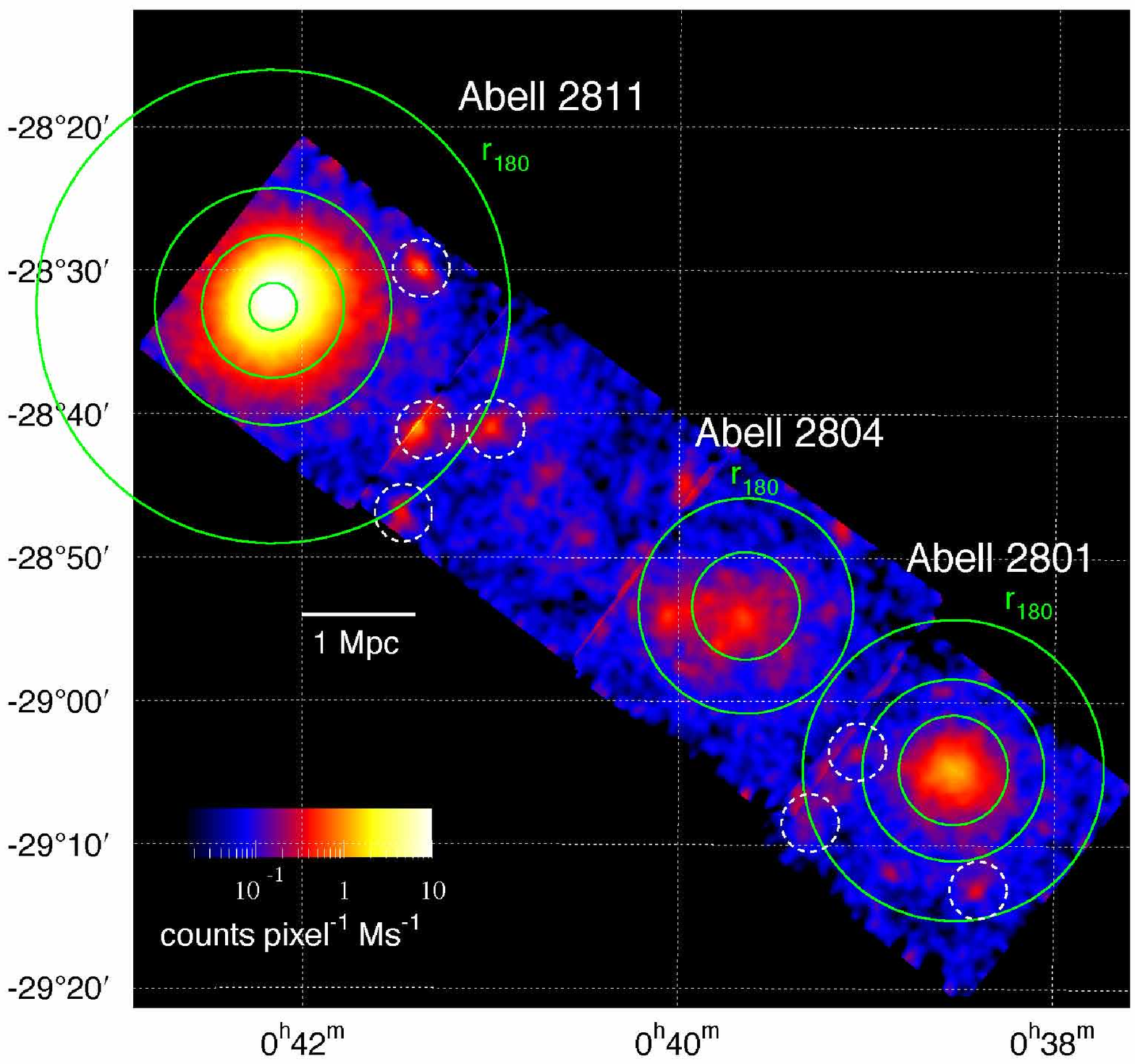}
\end{minipage}
\begin{minipage}{0.45\textwidth}
\FigureFile(\textwidth,\textwidth){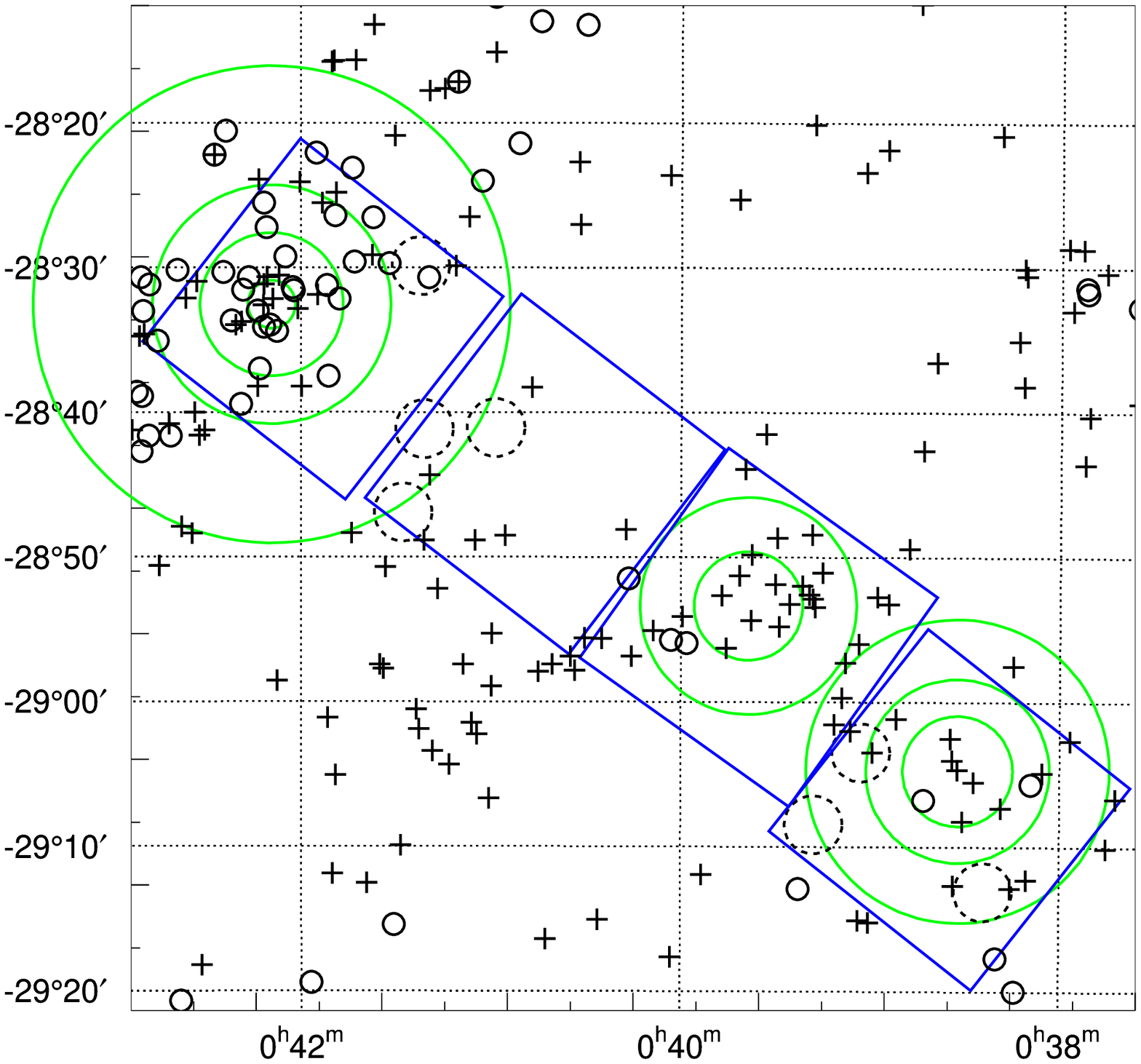}
\end{minipage}
\caption{Left: 
Combined XIS image of the four observations of the Sculptor supercluster 
in the 0.5--5.0 keV energy range. 
The observed XIS0, 1, 2, and 3 images 
were added on the sky coordinate,
and smoothed with $\sigma=16$ pixel $\simeq 17''$ Gaussian.
The instrumental background (NXB) were subtracted,
and the exposure was corrected, though vignetting was not corrected.
The green circles show the extracted region for the spectral analysis 
as shown in subsection \ref{subsec:spec}\@. The subtracted point source 
regions are indicated by the white dashed circles.
Right: 
Galaxy distribution associated with the Sculptor supercluster 
from 2dFGRS catalogue. The galaxies are selected between 
$0.103 < z < 0.113$, and the nearer ($z<0.108$) and farther ($z>0.108$) 
galaxies than the sculptor redshift correspond to the solid circles 
and crosses, respectively. The blue boxes show the observation regions 
with Suzaku, and the green circle and dashed circles are same as the 
left panel. 
}\label{fig:1}
\end{center}
\end{figure*}

A large part of baryons in the local universe is
thought to exist along the large-scale structure as filaments of the
intergalactic medium, called as warm-hot intergalactic medium
(WHIM), based on several N-body and gas dynamics simulations of
cosmological large-scale structure formation (e.g.\
\cite{cen99,cen06a,cen06b,dave01,chen03}).  The WHIM, whose
temperature and density is thought to be about $T=10^{5-7}$~K and
$10^{-6}$--$10^{-4}$~cm$^{-3}$, is hard to detect with the present-day
detectors such CCD cameras.  Thermal and chemical properties of the
WHIM, if observed, would bring us rich information about the structure
formation and evolution of the universe.

Recent Chandra and XMM-Newton observations with the high resolution
imaging spectroscopy enabled us to search for the WHIM emission
\citep{soltan02,kaastra03,finoguenov03,galeazzi07,
  takei07a,bregman07,werner08,galeazzi09}. \citet{werner08} and
\citet{galeazzi09} reported the significant detection of the emission
from the WHIM gas, while the others could not because of the extremely
low density of the gas compared with the high level of the the
Galactic foreground emission and the intrinsic background.  On the
other hand, Chandra and XMM observations with grating spectrometers
also made the absorption-line study of the WHIM possible and
restricted the WHIM density more significantly
\citep{fang02,fujimoto04,nicastro05,kaastra06,fang07,rasmussen07,takei07a,
  buote09,fang10,zappacosta10}.

Because Suzaku XIS is characterized by lower background level and
higher sensitivity below 1 keV \citep{koyama07}, we can observe the
intra-cluster medium (ICM) emission to the outer region of clusters
near the virial radii
\citep{sato07a,sato07b,bautz09,reiprich09,hoshino10}.  Although
\citet{takei07b} detected no significant redshifted O lines from the
outer region of A2218 with Suzaku, they set a tight constraint on
the intensity (see also \cite{tamura08}).  Reliable estimation of the
foreground Galactic emission is most important in constraining the
WHIM emission. Past studies have shown that Suzaku XIS can effectively
constrain the Galactic emission \citep{gupta09,yoshino09} better than
the previous satellites.

The Sculptor supercluster \citep{schuecker91,seitter92} at a redshift
of $z=0.11$ is one of the richest supercluster, composed of more than
20 clusters of galaxies \citep{einasto97} including 5 Abell
systems. It is located near the south galactic pole. Since WHIM is
thought to distribute along the filamentary structures over several
tens of Mpc scale connecting clusters of galaxies, it is likely that
superclusters contain significant amount of WHIM\@.  The Sculptor
supercluster has been observed with ROSAT and
ASCA\@. \citet{spiekermann96} and \citet{obayashi98} found no
indications of a largely extended X-ray emission in their observed
regions, which is overlapping with our Suzaku field. On the other
hand, \citet{zappacosta05} reported extended soft X-ray emission in
0.1 --0.3 keV band with ROSAT PSPC, and the region corresponds to
$\sim2$--4 degrees east of our observations. They suggested that the
diffuse emission was tracing the WHIM\@.

\citet{fang10} reported a $3\sigma$ detection of an oxygen absorption
line from the WHIM in the ``Sculptor Wall''at $z\sim0.03$ using the
Chandra and XMM grating observations of a blazer behind this
structure.  The Sculptor Wall is located much closer than the Sculptor
supercluster in redshift and $\sim 10$ degrees westward from our
observations. The estimated hydrogen density is $n=6\times10^{-6}$
cm$^{-3}$ under the assumption of the uniform absorber distribution
along the line of sight between $0.028<z<0.032$ using the
O\emissiontype{VII} column density.  \citet{zappacosta10} also
reported results from the WHIM search in ``Pisces--Cetus
Supercluster'' at $z=0.062$ and ``Farther Sculptor Wall'' at $z=0.128$
in the same line of sight as ``Sculptor Wall''.  In contrast to the
result of ``Sculptor Wall'', significant absorptions ($\sigma>3$) were
not detected from any of the strong H-like and He-like transitions of
C, O, and Ne at the redshift.

We use $H_0=70$ km~s$^{-1}$~Mpc$^{-1}$, $\Omega_{\Lambda} = 1-\Omega_M
= 0.73$ in this paper.  The redshifts of individual clusters are
0.1079 for A2811 and the offset observations, 0.108 for A2801 and
A2804, respectively, taken from NED as shown table \ref{tab:1}.  
At a redshift of $z=0.108$,
$1'$ corresponds to 119~kpc, and the virial radii, $r_{\rm 180} =
1.95\; h_{100}^{-1}\sqrt{k\langle T\rangle/10~{\rm keV}}$~Mpc
\citep{markevitch98}, is 1.96~Mpc ($16.5'$) for an average temperature
of $k\langle T\rangle = 5.0$~keV of A2811, 0.88~Mpc ($7.3'$) for
$k\langle T\rangle = 1.0$~keV of A2804, and 1.25~Mpc (10.5$'$) for
$k\langle T\rangle = 2.0$~keV of A2801.  Because of this redshift, we
can separate the WHIM O line from the supercluster relatively easily
from the Galactic emission.  Throughout this paper we adopt the
Galactic hydrogen column density of $N_{\rm H} = 1.54, 1.78,
1.70\times 10^{20}$ cm$^{-2}$ \citep{dickey90} in the direction of
A2811, A2801, A2804, respectively.  Because the system is located near
the south galactic pole, the Galactic absorption and emission are both
low and have no complex feature.  Unless noted otherwise, the solar
abundance table is given by \citet{anders89}, and the errors are in
the 90\% confidence region for a single interesting parameter.

\section{Observations and Data Reduction}
\subsection{Observations}
\label{subsec:obs}
Suzaku observed the Sculptor supercluster in November and December 2005.  
The observation logs are given in table~\ref{tab:2}, 
and the XIS image in 0.5--5.0~keV is shown in figure~\ref{fig:1} left.
Figure~\ref{fig:1} right shows a galaxy density map in $0.103<z<0.113$ 
using the 2dF Galaxy Redshift Survey (2dFGRS) 
\footnote{http://www.mso.anu.edu.au/2dFGRS/}.
The XIS was operated in the Normal clocking mode (8~s exposure 
per frame), with the standard $5\times 5$ or $3\times 3$ editing mode.
During these observations, the significant effect of 
the Solar Wind Charge eXchange (SWCX) was not confirmed in ACE data
\footnote{http://www.srl.caltech.edu/ACE/ASC/},
while it was known that the SWCX affected some Suzaku spectra in the 
lower energy range as reported in \citet{fujimoto07}, \citet{gupta09},
\citet{koutroumpa09} and \citet{yoshino09}.

\subsection{Data Reduction}
\label{subsec:data}
We used version 2.1 processing data, and the analysis was performed
with HEAsoft version 6.6.3 and XSPEC 12.5.0ac.  In the analysis of XIS
data, we selected ELEVATION $>$ 15$^\circ$ of the data set to remove
stray-light from the day Earth limb.  Event screening with cut-off
rigidity (COR) was not performed in our data. Note
that, even if we select the data with COR$>$8 GV, the resultant fit
parameters all stay within the statistical errors.
In order to subtract the non X-ray background (NXB), 
we employed the dark Earth database by the ``xisnxbgen'' Ftools task.

For the spectral fits of the ICM emission, we generated 
Ancillary Response Files (ARFs) for A2801/A2811 assumed 
the $\beta$-model surface brightness profile, 
$\beta = 0.8/0.82$ and $r_c =$ \timeform{0.88'}/\timeform{1.6'}, 
respectively, reported in \citet{obayashi98}, 
by the ``xissimarfgen'' Ftools task \citep{ishisaki07}.
We also generated the ARFs for each observation
assumed an uniform sky of 20$'~\times~20'$ for the Galactic 
and Cosmic X-ray Background emission. 
Note that we used the uniform ARFs for the spectral fits of A2804. 
We included the effect of the contaminations on the optical blocking 
filter of the XISs in the ARFs.
Since the energy resolution also slowly degraded
after the launch, due to radiation damage, this effect was included
in the Redistribution Matrix File (RMF) by the ``xisrmfgen'' Ftools task.

\section{Results and Discussion}

\subsection{Strategy of the spectral fits}
\label{subsec:strategy}

We analyzed the spectra of the Sculptor supercluster using the following way.

\begin{figure*}[ht]
\begin{minipage}{0.4\textwidth}
\FigureFile(\textwidth,\textwidth){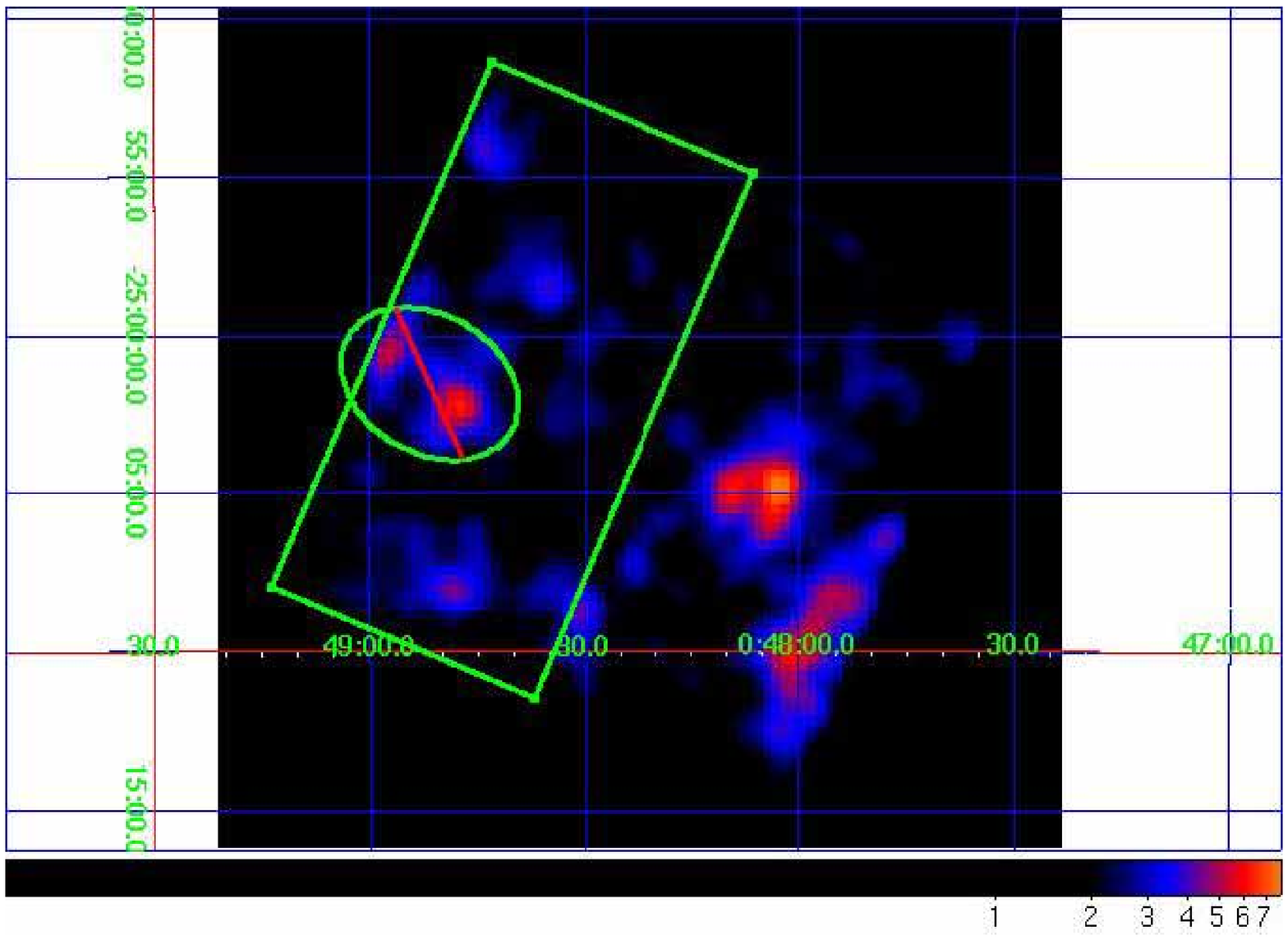}
\end{minipage}\hfill
\begin{minipage}{0.4\textwidth}
\FigureFile(\textwidth,\textwidth){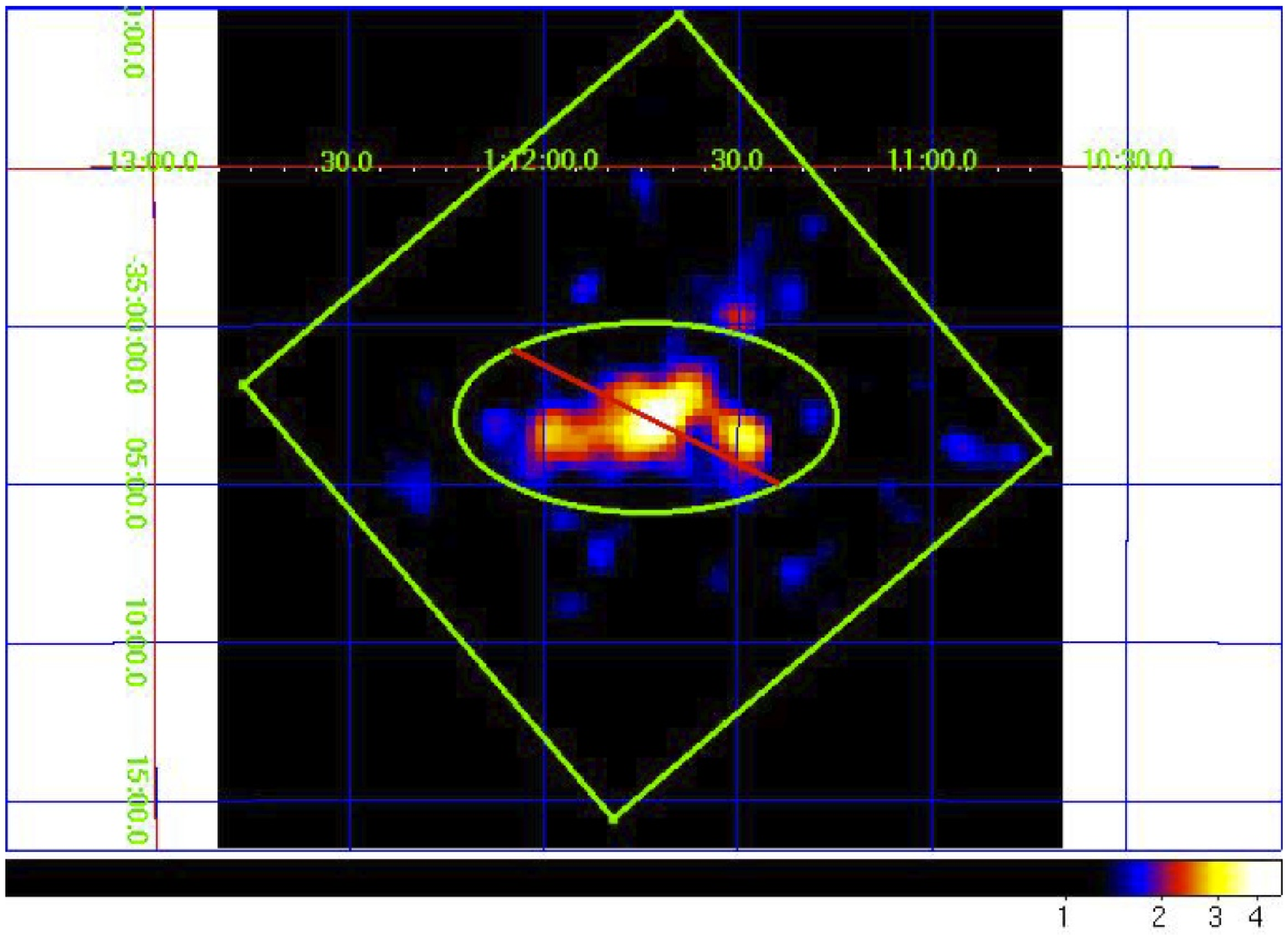}
\end{minipage}
\caption{
Extracted regions of NGC~253 offset (Left) and Q0109-3518 (Right). 
Green boxes corresponds to the extracted region for the spectral fits, 
and the elliptical regions are eliminated from the spectral region.  
}\label{fig:3}
\end{figure*}

\begin{enumerate}

\item Extraction of the point-like sources for each observation.

\item Background estimation for the Galactic and Cosmic X-ray background
      (CXB).

     \begin{enumerate}
     
     \item Spectral fits of the outer region of the virial radii 
           of the member clusters in A2811 offset region as shown
           in figure~\ref{fig:1}.

     \item Estimation of the background models using the Suzaku observations 
           near the Sculptor supercluster as shown in figure~\ref{fig:3}.

     \item Comparing the above two results, we determine the background model.

     \end{enumerate}

\item Spectral analysis of the ICM emission of A2801, A2804, A2811 to the
      virial radius.

\item Constraining the O emissions from WHIM in the filament region between the clusters.

\end{enumerate}
\begin{figure*}
\begin{minipage}{0.24\textwidth}
\FigureFile(\textwidth,\textwidth){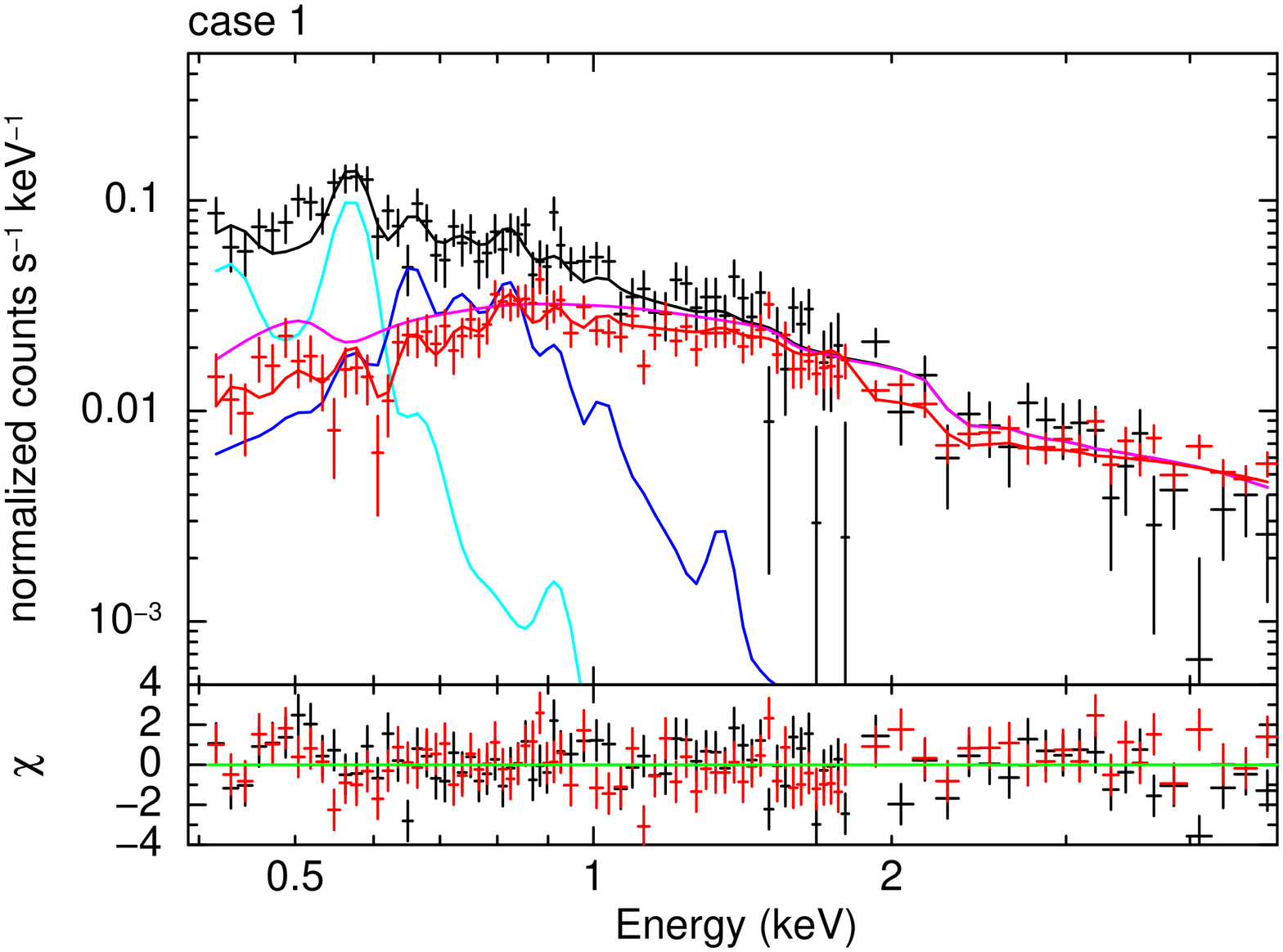}
\end{minipage}
\begin{minipage}{0.24\textwidth}
\FigureFile(\textwidth,\textwidth){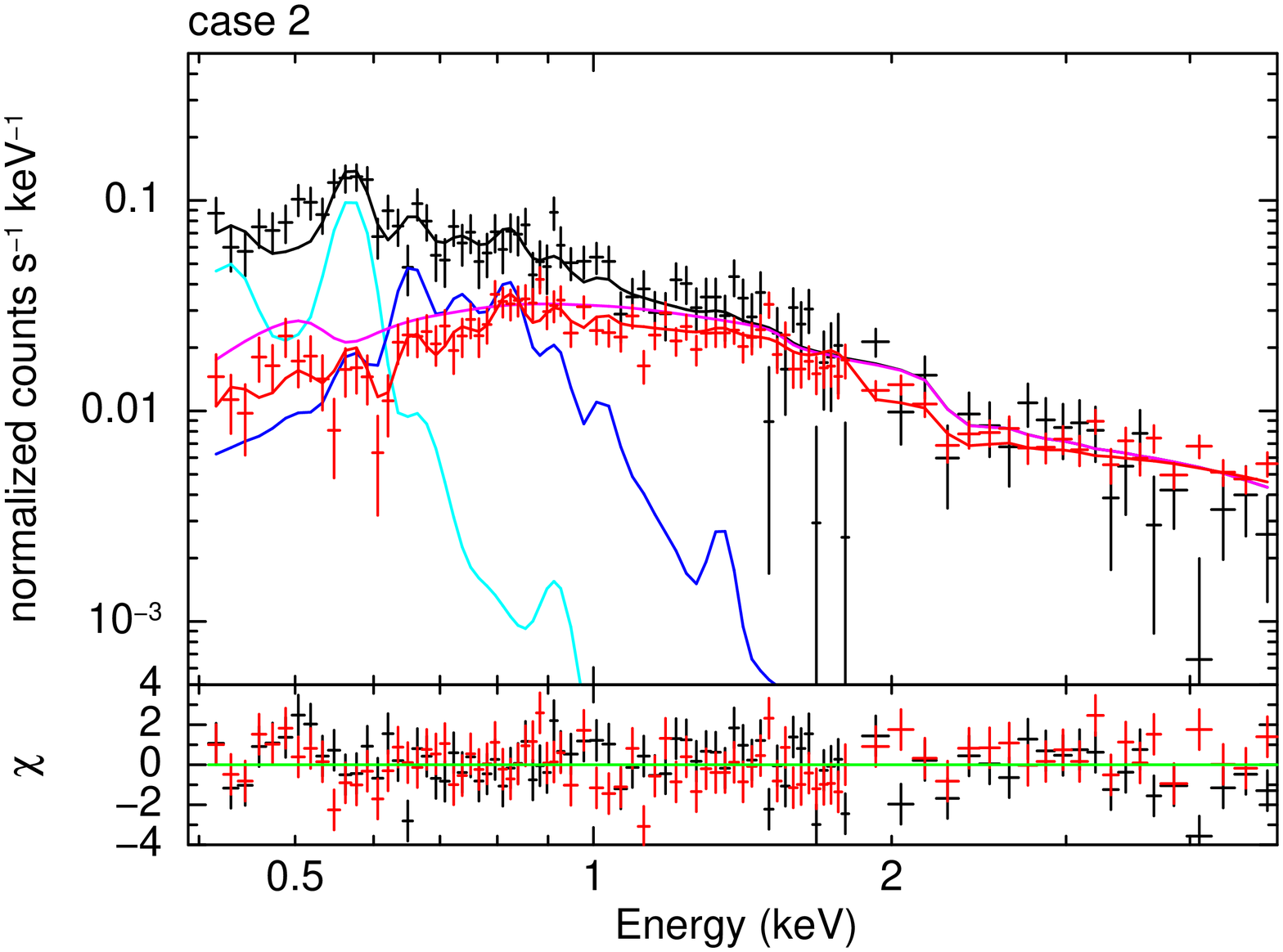}
\end{minipage}
\begin{minipage}{0.24\textwidth}
\FigureFile(\textwidth,\textwidth){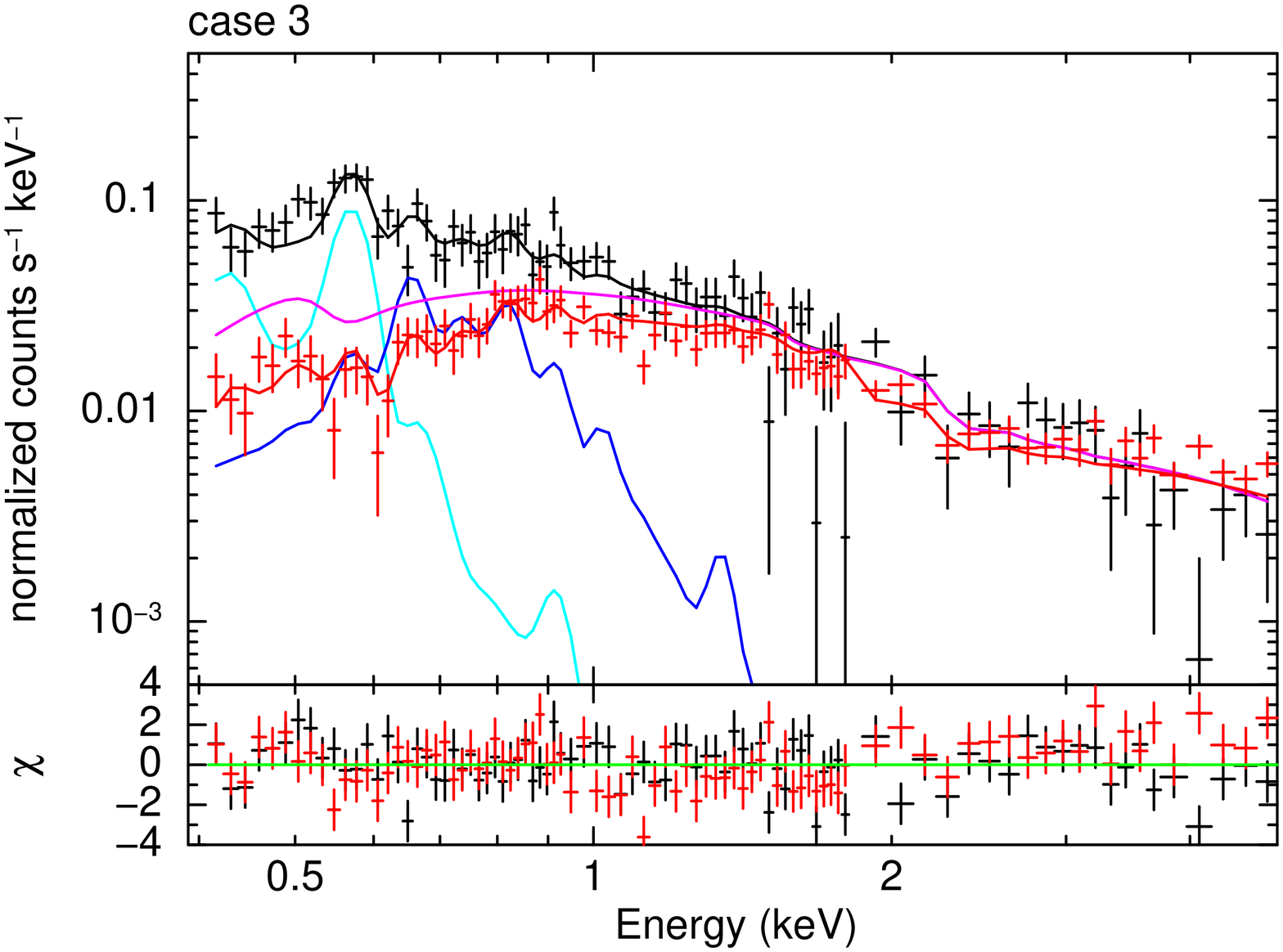}
\end{minipage}
\begin{minipage}{0.24\textwidth}
\FigureFile(\textwidth,\textwidth){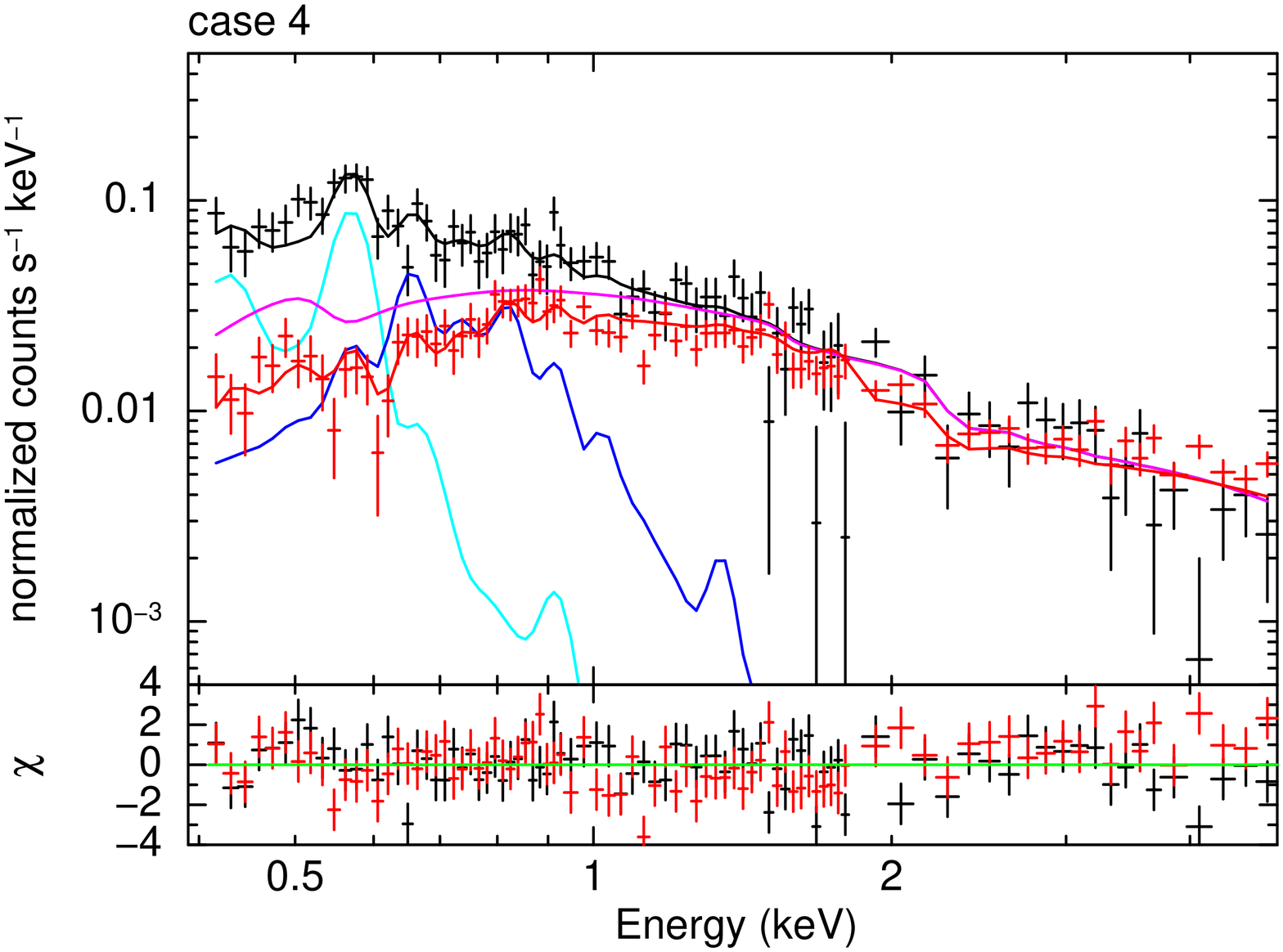}
\end{minipage}

\begin{minipage}{0.24\textwidth}
\FigureFile(\textwidth,\textwidth){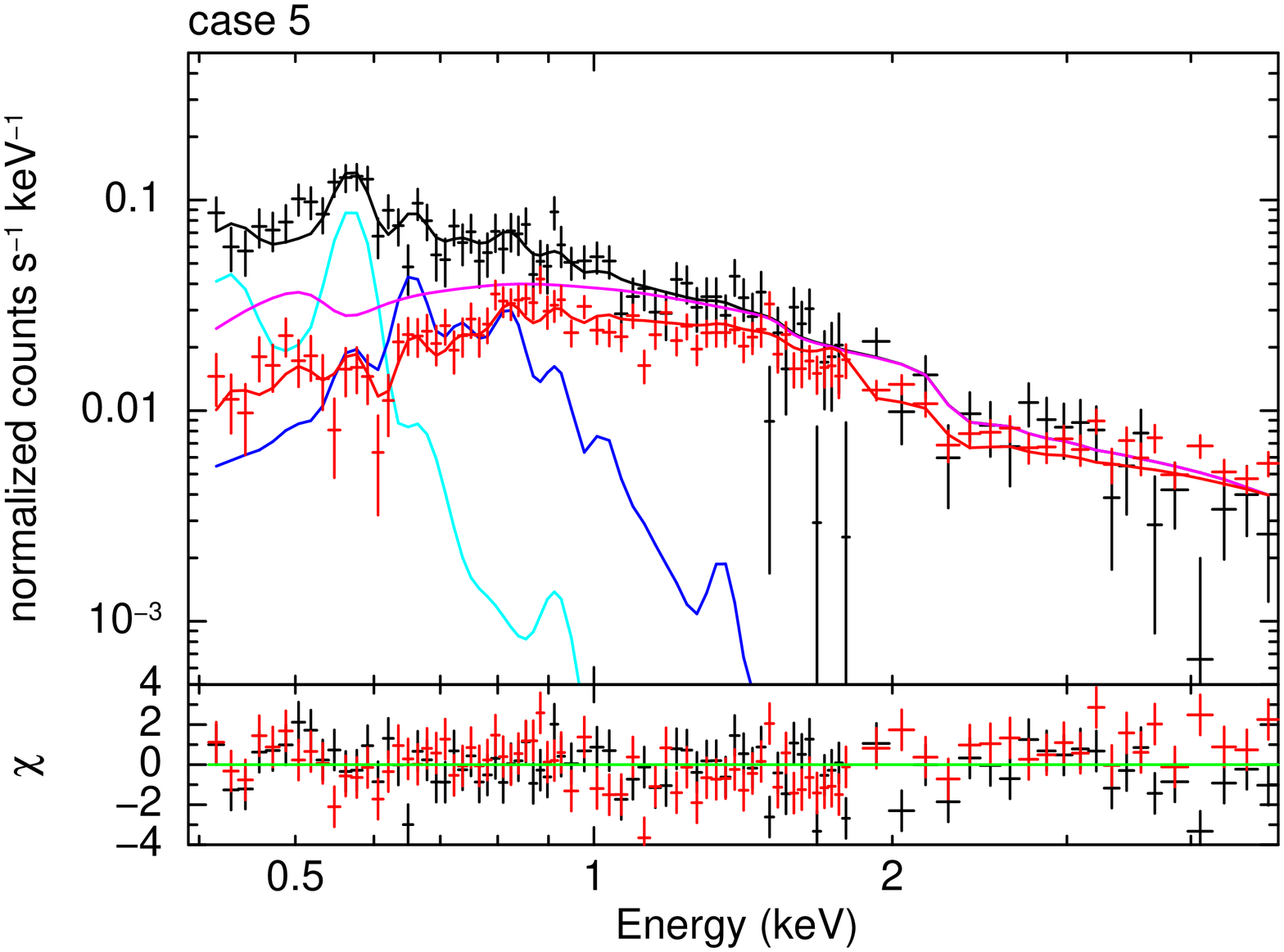}
\end{minipage}
\begin{minipage}{0.71\textwidth}
\caption{
The panels show the resultant fits of the outer region of the virial 
radii of the member clusters 
with the models as shown in table \ref{tab:3}. 
The data are plotted by black and red crosses for BI and FI, 
and the black and red lines show the best-fit model for the BI and 
FI spectra, respectively. The CXB components are shown in a 
magenta line, and the Galactic emissions are indicated by cyan and 
blue lines, respectively.}
\label{fig:2}
\end{minipage}
\end{figure*}
\begin{table*}
\caption{Summary of the parameters of the fits of the outer region of
 the virial radii of the clusters. We assumed zero redshift and a solar 
metallicity in the fits.}
\label{tab:3}
\begin{center}
\begin{tabular}{lccccc} \hline 
 & case~1 & case~2 & case~3 & case~4 & case~5\\
\hline
LHB & & & & & \\
$kT$ (keV) & 0.10$^{+0.01}_{-0.01}$ & 0.10(fixed) & $\leftarrow$ (fixed)
	     & $\leftarrow$ (fixed) &  $\leftarrow$ (fixed)\\
$Norm^{\ast}$ ($\times10^{-3}$) & 2.62$^{+1.57}_{-1.00}$ &
	 2.61$^{+0.37}_{-0.37}$& 2.36$^{+0.36}_{-0.36}$  & 2.32$^{+0.31}_{-0.31}$ & 2.32$^{+0.32}_{-0.32}$ \\
\hline
MWH & & & & & \\
$kT$ (keV) & 0.32$^{+0.03}_{-0.03}$ & 0.32$^{+0.03}_{-0.02}$ & 0.31$^{+0.03}_{-0.03}$  & 0.30(fixed) & $\leftarrow$ (fixed)\\
$Norm^{\ast}$ ($\times10^{-4}$) & 4.18$^{+0.71}_{-0.39}$ 
& 4.18$^{+0.67}_{-0.67}$ & 3.60$^{+0.63}_{-0.63}$ & 3.69$^{+0.50}_{-0.50}$ & 3.54$^{+0.51}_{-0.50}$ \\
\hline
CXB & & & & & \\
$\Gamma$ & 1.22$^{+0.08}_{-0.06}$ & 1.22$^{+0.08}_{-0.08}$  & 1.4(fixed)
	     & $\leftarrow$ (fixed) & $\leftarrow$ (fixed) \\
$Norm^{\ast,~\dagger}$ ($\times10^{-4}$) & 7.93$^{+0.64}_{-0.62}$ 
& 7.93$^{+0.65}_{-0.62}$  & 8.99$^{+0.48}_{-0.47}$  &  9.01$^{+0.48}_{-0.47}$ & 9.6 (fixed) \\
\hline
 $\chi^2/\nu$ & 218/167 & 218/168 & 231/169 & 231/170 & 235/171 \\
\hline\\[-1ex]
\multicolumn{6}{l}{\parbox{0.7\textwidth}{\footnotesize 
\footnotemark[$\ast$] 
Normalization of the {\it apec} component
divided by the solid angle, $\Omega^{\makebox{\tiny\sc u}}$,
assumed in the uniform-sky ARF calculation (20$'$ radius),
${\it Norm} = \int n_{\rm e} n_{\rm H} dV \,/\,
(4\pi\, (1+z)^2 D_{\rm A}^{\,2}) \,/\, \Omega^{\makebox{\tiny\sc u}}$
$\times 10^{-14}$ cm$^{-5}$~400$\pi$~arcmin$^{-2}$, 
where $D_{\rm A}$ is the angular distance to the source.}}\\
\multicolumn{6}{l}{\parbox{0.6\textwidth}{\footnotesize 
\footnotemark[$\dagger$]
The referred CXB normalization in \citet{kushino02} is 9.6$\times10^{-4}$
for $\Gamma=1.4$ in the above unit. }}\\
\end{tabular}
\end{center}
\end{table*}
\begin{table*}
\caption{List of $\chi^2$/dof in case~4 in table~\ref{tab:3} 
for the fits of the nominal and
considering the systematic errors such as 
the NXB level and contaminant of OBF. For details, see text.}
\label{tab:4}
\begin{center}
\begin{tabular}{lccccc}
\hline\hline
\makebox[6em][l]{uncertainties} & nominal &\multicolumn{2}{c}{NXB} & \multicolumn{2}{c}{contaminant}\\
\hline
 & & \makebox[0in][c]{+10\%} & \makebox[0in][c]{-10\%} & \makebox[0in][c]{+10\%} & \makebox[0in][c]{-10\%}\\
\hline
$\chi^2$/dof $\dotfill$   & 231/170 & 247/170 & 227/170 & 231/170 & 231/170 \\
\hline
\end{tabular}
\end{center}
\end{table*}

\begin{table*}[ht]
\caption{Suzaku Observation logs of NGC~253 offset and Q0109-3518.}
\label{tab:5}
\begin{tabular}{lcccccc} \hline 
Region & Seq. No. & Obs. date & \multicolumn{1}{c}{(RA, Dec)$^\ast$}
 &Exp.&After screening & Distance$^\dagger$ \\
&&&J2000& ksec &(BI/FI) ksec & degree \\
\hline 
NGC~253 offset & 803004010 & 2008-12-29 & (\timeform{00h48m22.2s},
 \timeform{-25D02'51''})& 56.7 & 53.7/53.7 & 3.75\\
Q0109-3518 & 703037010 & 2008-05-20 & (\timeform{01h11m45.0s},
 \timeform{-35D03'10''})& 30.0 & 26.0/25.9 & 9.01\\
\hline\\[-1ex]
\multicolumn{7}{l}{\parbox{0.9\textwidth}{\footnotesize 
\footnotemark[$\ast$]
Average pointing direction of the XIS, written in the 
RA\_NOM and DEC\_NOM keywords of the event FITS files.}}\\
\multicolumn{7}{l}{\parbox{0.9\textwidth}{\footnotesize 
\footnotemark[$\dagger$]
Distance from  A2811}}\\
\end{tabular}
\end{table*}
\begin{figure*}[!ht]
\begin{minipage}{0.24\textwidth}
\FigureFile(\textwidth,\textwidth){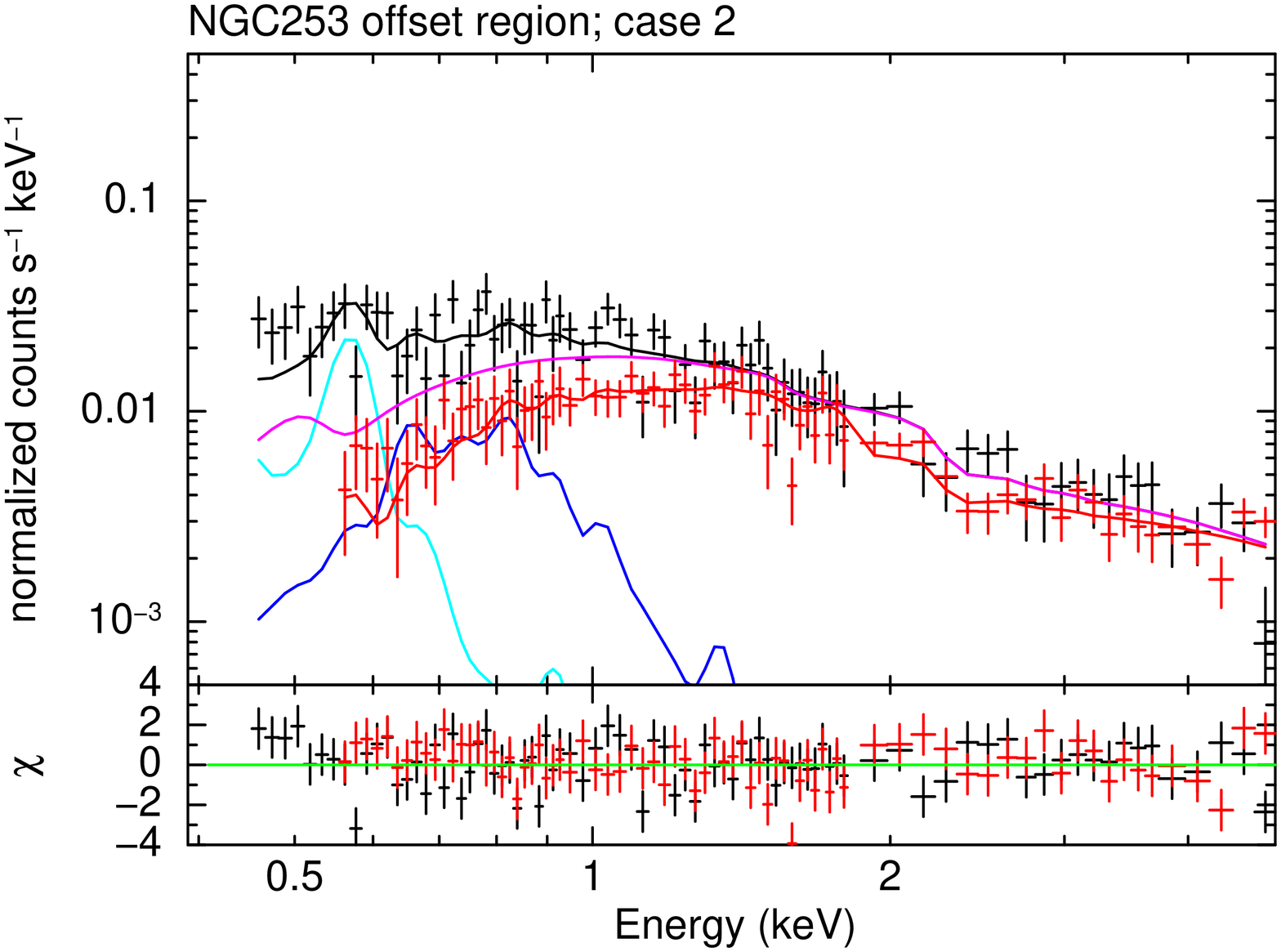}
\end{minipage}
\begin{minipage}{0.24\textwidth}
\FigureFile(\textwidth,\textwidth){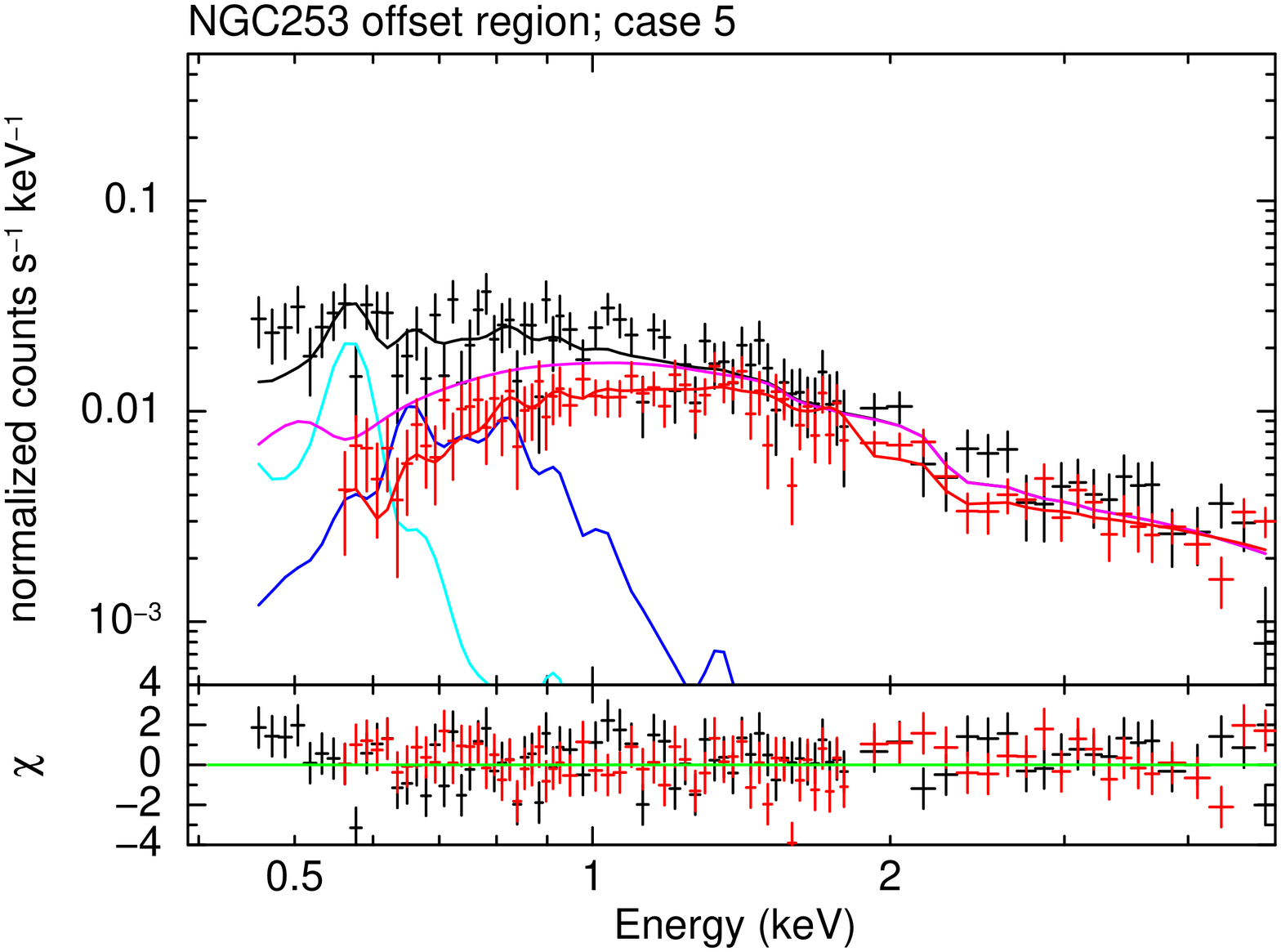}
\end{minipage}
\begin{minipage}{0.24\textwidth}
\FigureFile(\textwidth,\textwidth){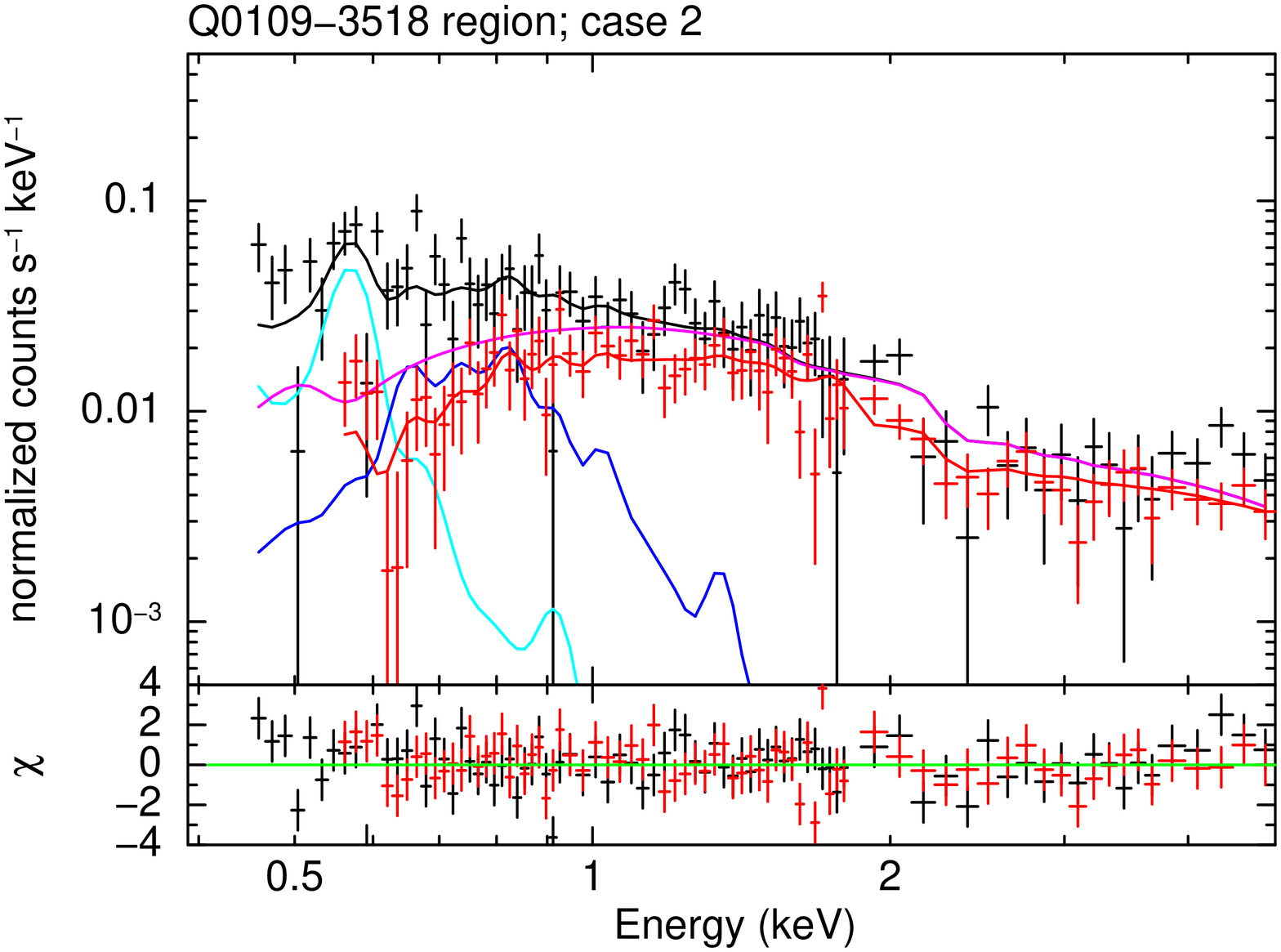}
\end{minipage}
\begin{minipage}{0.24\textwidth}
\FigureFile(\textwidth,\textwidth){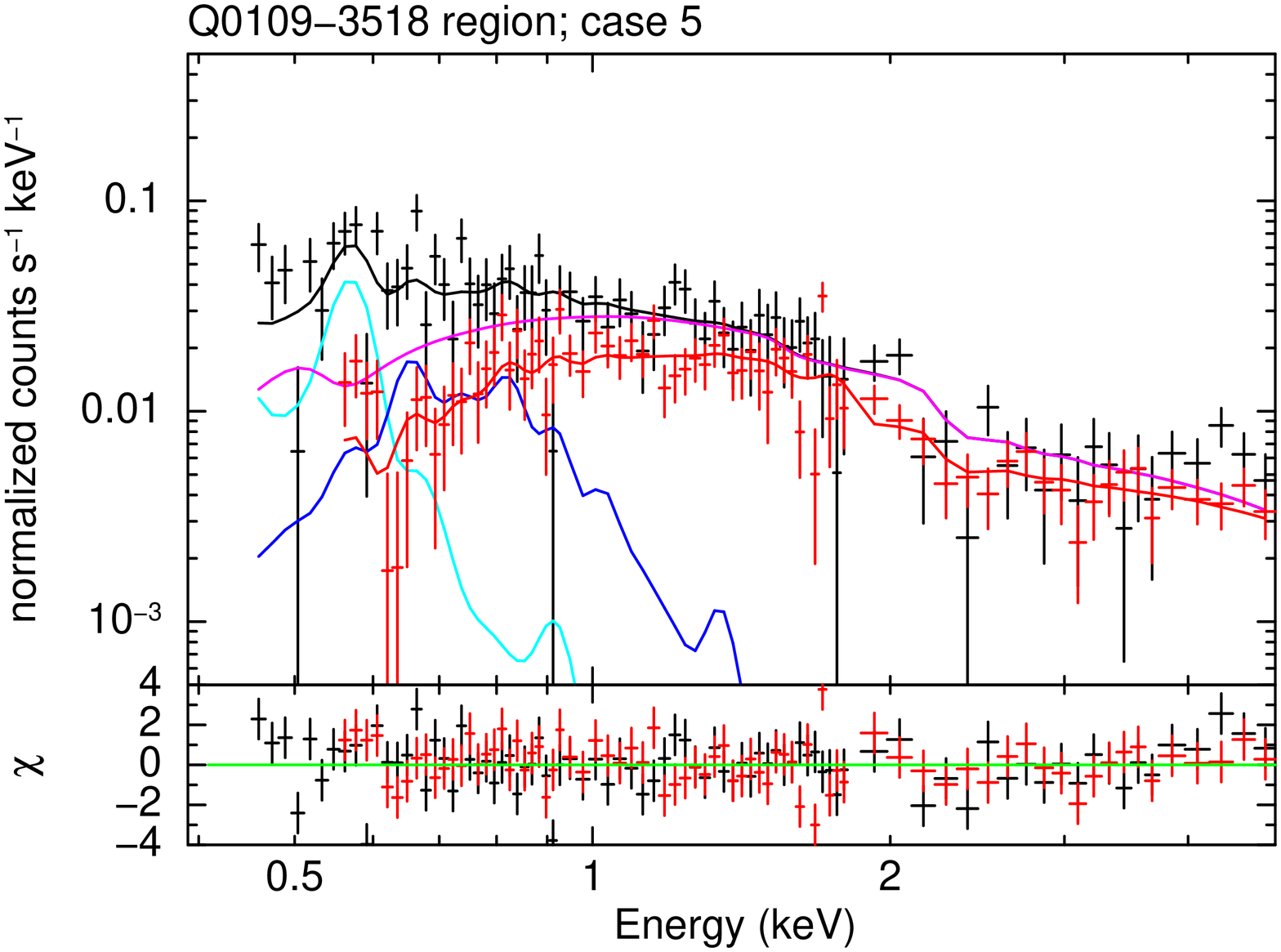}
\end{minipage}
\caption{
The spectra of NGC~253 offset and Q0109-3518\@. The notations 
in the figure are same as figure \ref{fig:2}.
}\label{fig:4}
\end{figure*}
\begin{table*}[!ht]
\caption{Summary of the parameters of the fits of the NGC~253 offset
 region and Q0109-3518 region\@.}
\label{tab:6}
\begin{center}
\begin{tabular}{lcccc} \hline 
Region & \multicolumn{2}{c}{NGC~253 offset} &
 \multicolumn{2}{c}{Q0109-3518} \\
\hline
$N_{\rm H}$ (cm$^{-2}$) &   \multicolumn{2}{c}{1.42$\times10^{20}$} &
 \multicolumn{2}{c}{1.93$\times10^{20}$}  \\
\hline
 & case~2 & case~5 & case~2 & case~5 \\
\hline
LHB & & & & \\
$kT$ (keV) & 0.10(fixed) & $\leftarrow$ (fixed)
	     & $\leftarrow$ (fixed)& $\leftarrow$ (fixed) \\
$Norm^{\ast}$ ($\times10^{-3}$) & 2.45$^{+0.74}_{-0.78}$ &
	 2.56$^{+0.65}_{-0.66}$ & 2.93$^{+0.76}_{-0.86}$ &
		2.57$^{+0.72}_{-0.72}$ \\
\hline
MWH & & & & \\
$kT$ (keV) & 0.32$^{+0.08}_{-0.06}$ & 0.3 (fixed)
	     & 0.34$^{+0.07}_{-0.06}$ & 0.3 (fixed) \\
$Norm^{\ast}$ ($\times10^{-4}$) & 2.59$^{+1.14}_{-1.02}$ &
	 3.00$^{+0.72}_{-0.71}$ & 2.98$^{+1.28}_{-1.17}$ &
		 2.79$^{+0.84}_{-0.83}$ \\
\hline
CXB & & & & \\
$\Gamma$ & 1.38$^{+0.09}_{-0.09}$ & 1.4 (fixed) &
		1.30$^{+0.13}_{-0.12}$  & 1.4 (fixed) \\
$Norm^{\ast,~\dagger}$ ($\times10^{-4}$) & 10.26$^{+0.94}_{-0.89}$ &
	 9.6 (fixed) & 8.49$^{+1.04}_{-0.98}$ & 9.6 (fixed) \\
\hline
 $\chi^2/\nu$ & 180/155 & 185/158 & 204/155 & 207/158 \\
\hline\\[-1ex]
\multicolumn{5}{l}{\parbox{0.7\textwidth}{\footnotesize 
\footnotemark[$\ast$] 
Normalization of the {\it apec} component
divided by the solid angle, $\Omega^{\makebox{\tiny\sc u}}$,
assumed in the uniform-sky ARF calculation (20$'$ radius),
${\it Norm} = \int n_{\rm e} n_{\rm H} dV \,/\,
(4\pi\, (1+z)^2 D_{\rm A}^{\,2}) \,/\, \Omega^{\makebox{\tiny\sc u}}$
$\times 10^{-14}$ cm$^{-5}$~400$\pi$~arcmin$^{-2}$, 
where $D_{\rm A}$ is the angular distance to the source.}}\\
\multicolumn{5}{l}{\parbox{0.6\textwidth}{\footnotesize 
\footnotemark[$\dagger$]
The referred CXB normalization in \citet{kushino02} is 9.6$\times10^{-4}$
for $\Gamma=1.4$ in the above unit. }}\\
\end{tabular}
\end{center}
\end{table*}

\subsection{Fore- and Background Estimation}
\label{subsec:bgd}
Using "wavdetect" tool in
CIAO\footnote{http://cxc.harvard.edu/ciao/}, we searched for
point-like sources in the Suzaku images in the energy range 0.5-2.0
keV\@. As shown by the white dashed circles in figure \ref{fig:1},
we detected and subtracted 7 point-like sources using 2 arcmin
radius which corresponded to the HPD of the Suzaku XRT.  The flux
level of the sources were $>1.5\times10^{-12}$ erg cm$^{-2}$
s$^{-1}$ in 0.5-2.0 keV, and contribution from unresolved sources
was taken into account when we subtract the CXB spectrum.  After
subtracting the point sources, we fitted the spectra integrated over
the A2811 offset region excluding the virial region ($r<r_{180}$)
of A2811, where a few galaxies existed as shown in
figure~\ref{fig:1} right, with the following models including the
local hot bubble (LHB) and milky way halo (MWH) as the Galactic
foregrounds and the CXB as a background: apec$_{\rm LHB}+{\rm
  phabs}\times({\rm apec}_{\rm MWH}+{\rm pow}_{\rm CXB})$\@.  It was
suggested that SWCX made a significant contribution to unabsorbed
emission-line component (LHB) even in the case of no flare in the
solar-wind flux (e.g.\ \cite{yoshino09}).  However, since SWCX and
thermal plasma is hardly distinguished with a limited spectral
resolution of CCDs \citep{gupta09}, we model the sum of SWCX and LHB
as a single unabsorbed thermal plasma.  Here, we assumed zero redshift
and a solar metallicity.  We examined the spectral fits with the
model, by changing the parameters to be free or fixed.  We also
assumed that there was no intracluster and/or intergalactic emission
outside of the virial radii. We discuss possible intergalactic
emission in subsection \ref{subsec:limit}.  The resultant parameters
of the fits are shown in table \ref{tab:3} and figure \ref{fig:2}.  As
a result, the spectra were well-fitted by the sum of Galactic and CXB
components.  We also tested the inferred photon index, $\Gamma$, and
normalization of the CXB model in \citet{kushino02} after the
point-source subtraction. Our results gave slightly lower intensity
than the \citet{kushino02} level, with poor statistics above 2 keV\@.
Thus, we employed the parameters of the case~5 in table~\ref{tab:3},
derived with fixed spectral shape for the Galactic emission and fixed
shape and intensity for the CXB component, respectively. We performed
spectral fit for the clusters in the Sculptor supercluster with this
fixed background in the following analysis.

\begin{figure*}
\begin{minipage}{0.32\textwidth}
\FigureFile(\textwidth,\textwidth){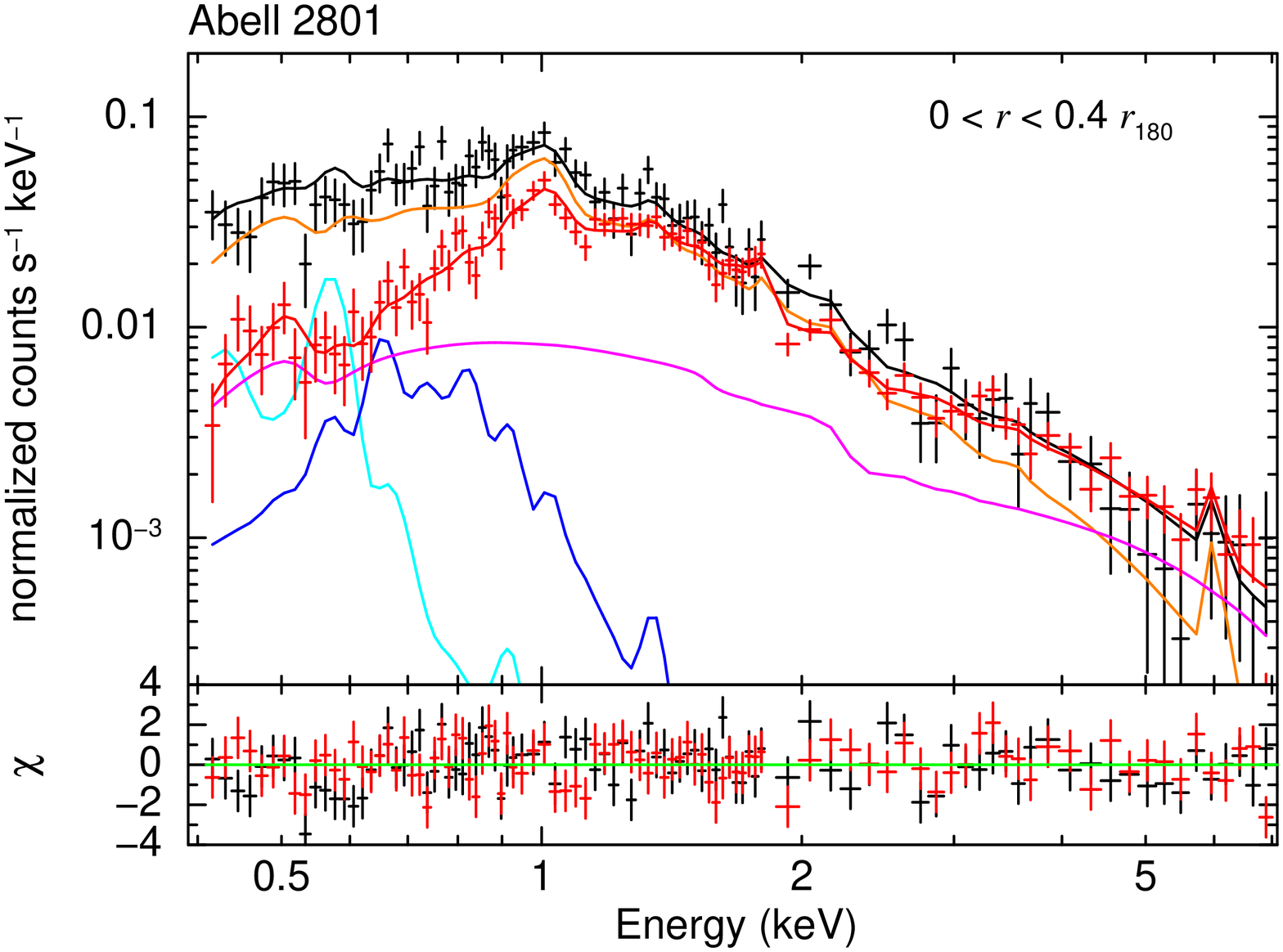}
\end{minipage}
\begin{minipage}{0.32\textwidth}
\FigureFile(\textwidth,\textwidth){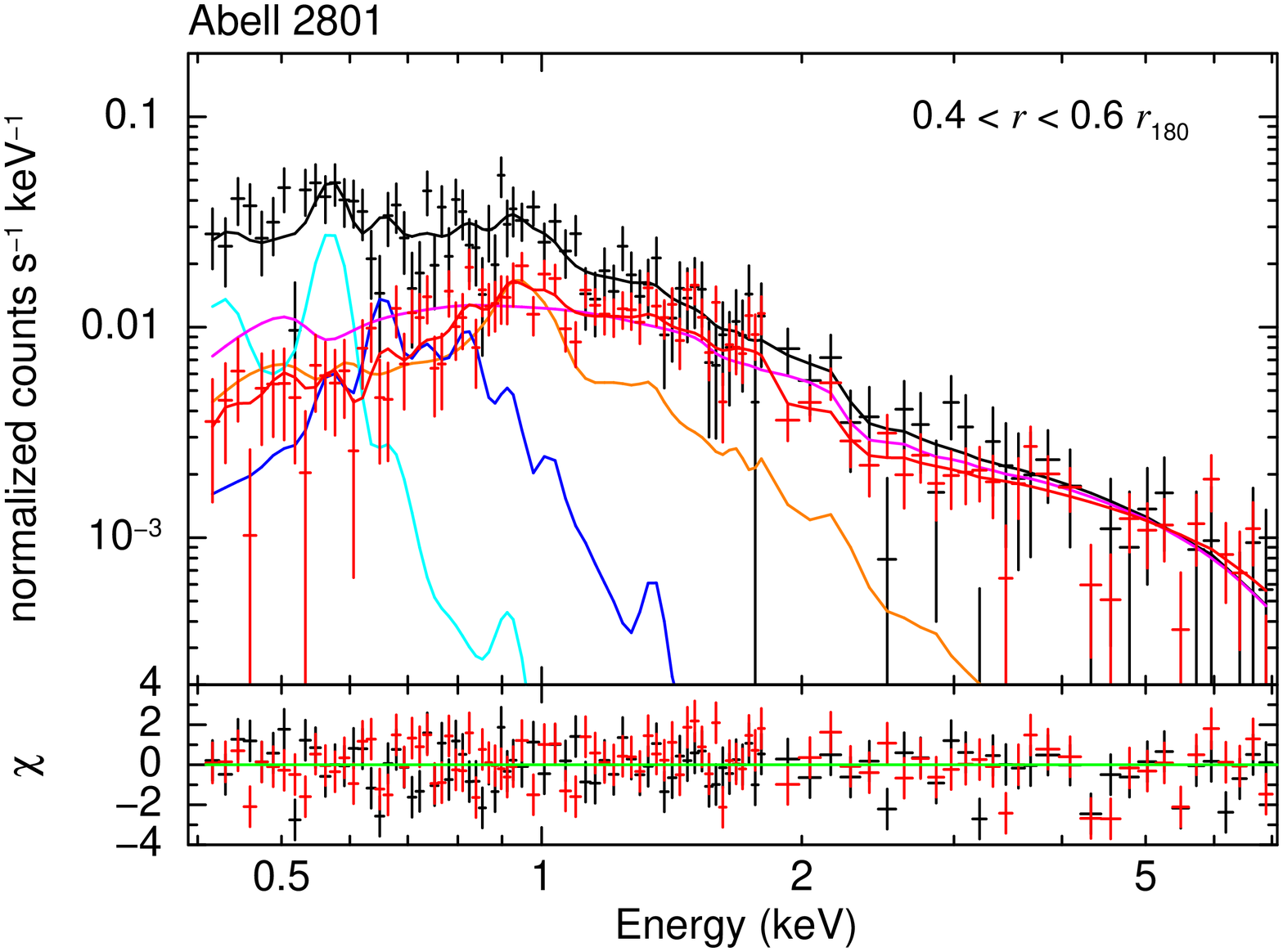}
\end{minipage}
\begin{minipage}{0.32\textwidth}
\FigureFile(\textwidth,\textwidth){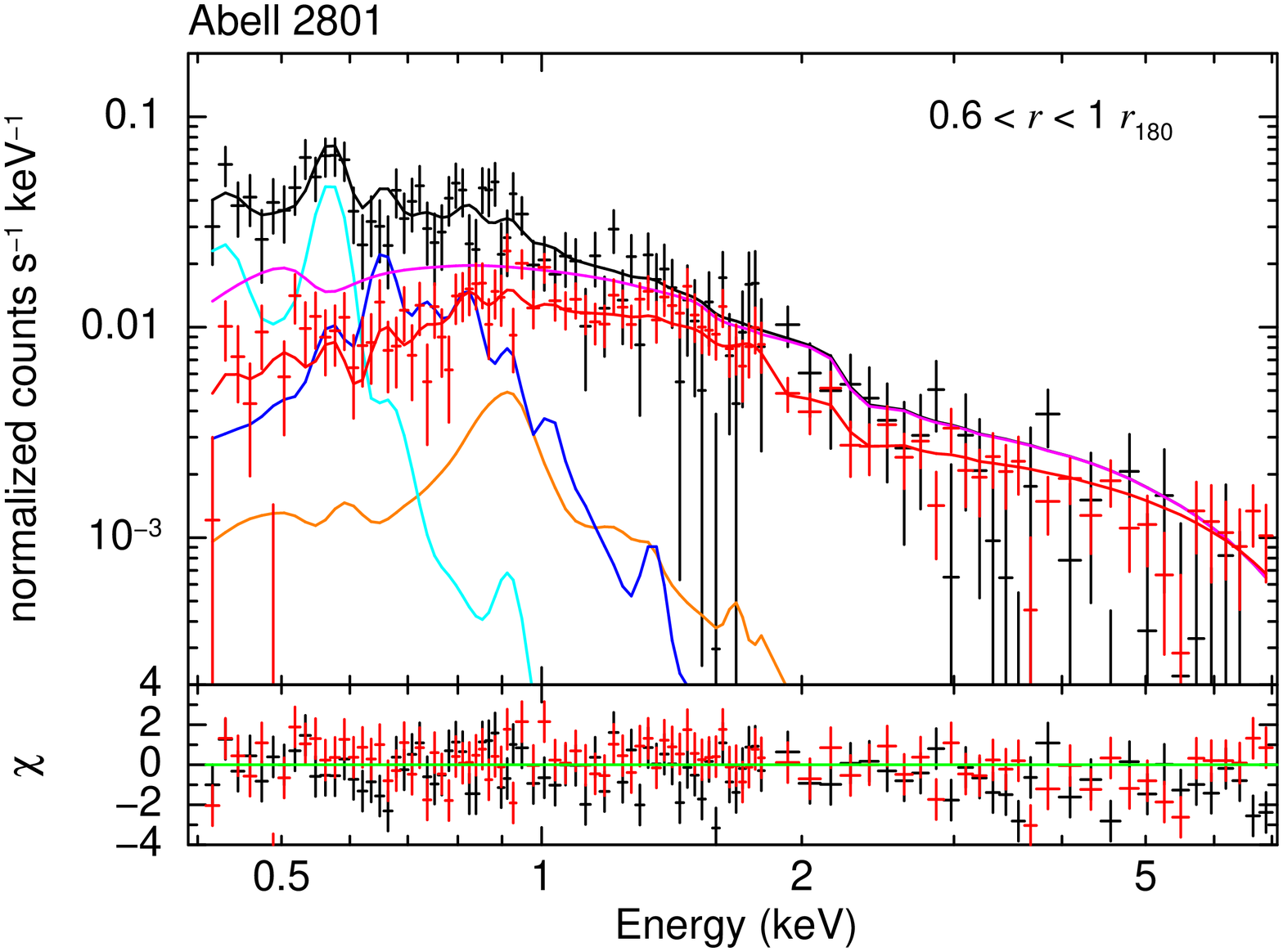}
\end{minipage}

\begin{minipage}{0.32\textwidth}
\FigureFile(\textwidth,\textwidth){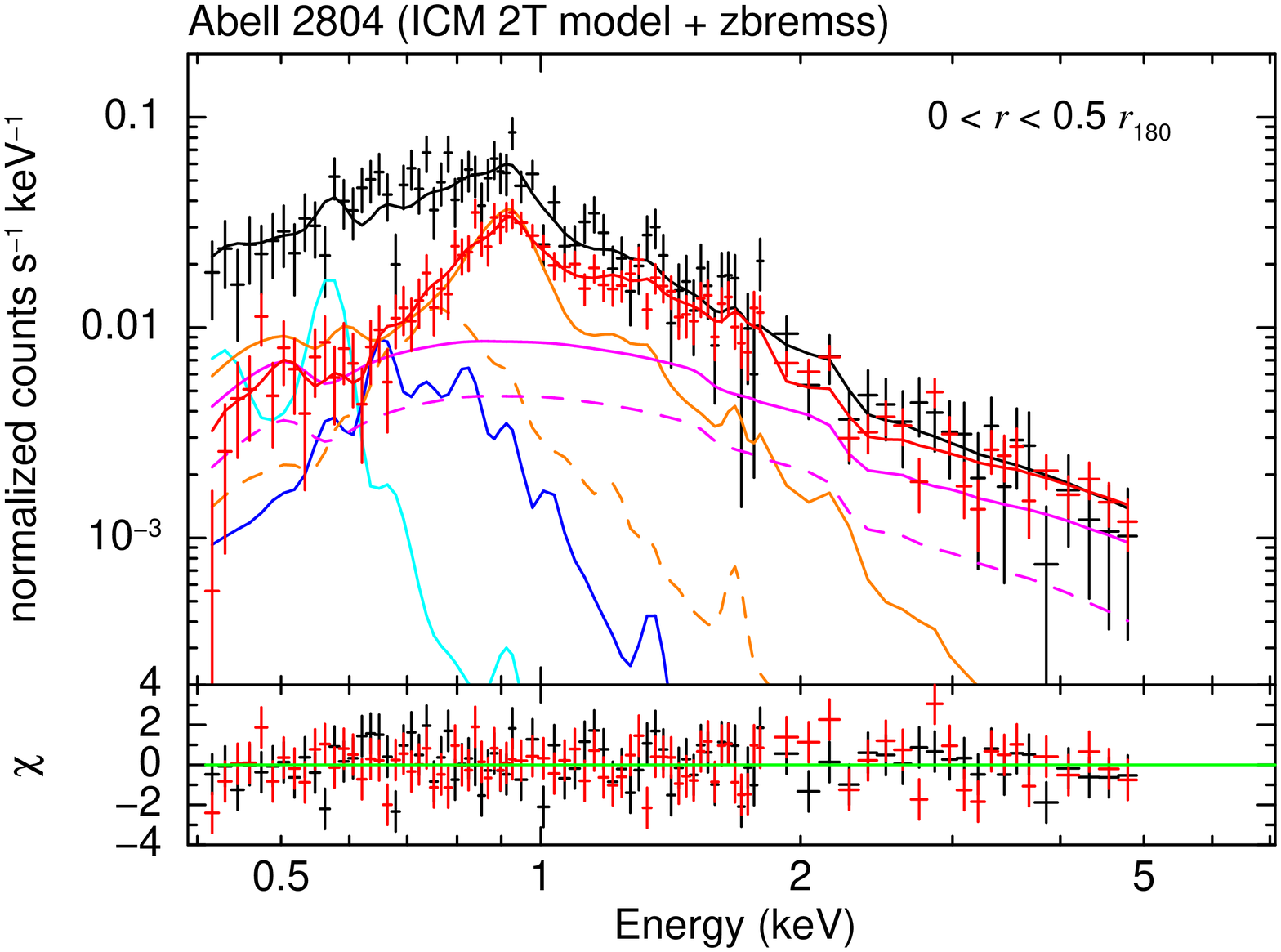}
\end{minipage}
\begin{minipage}{0.32\textwidth}
\FigureFile(\textwidth,\textwidth){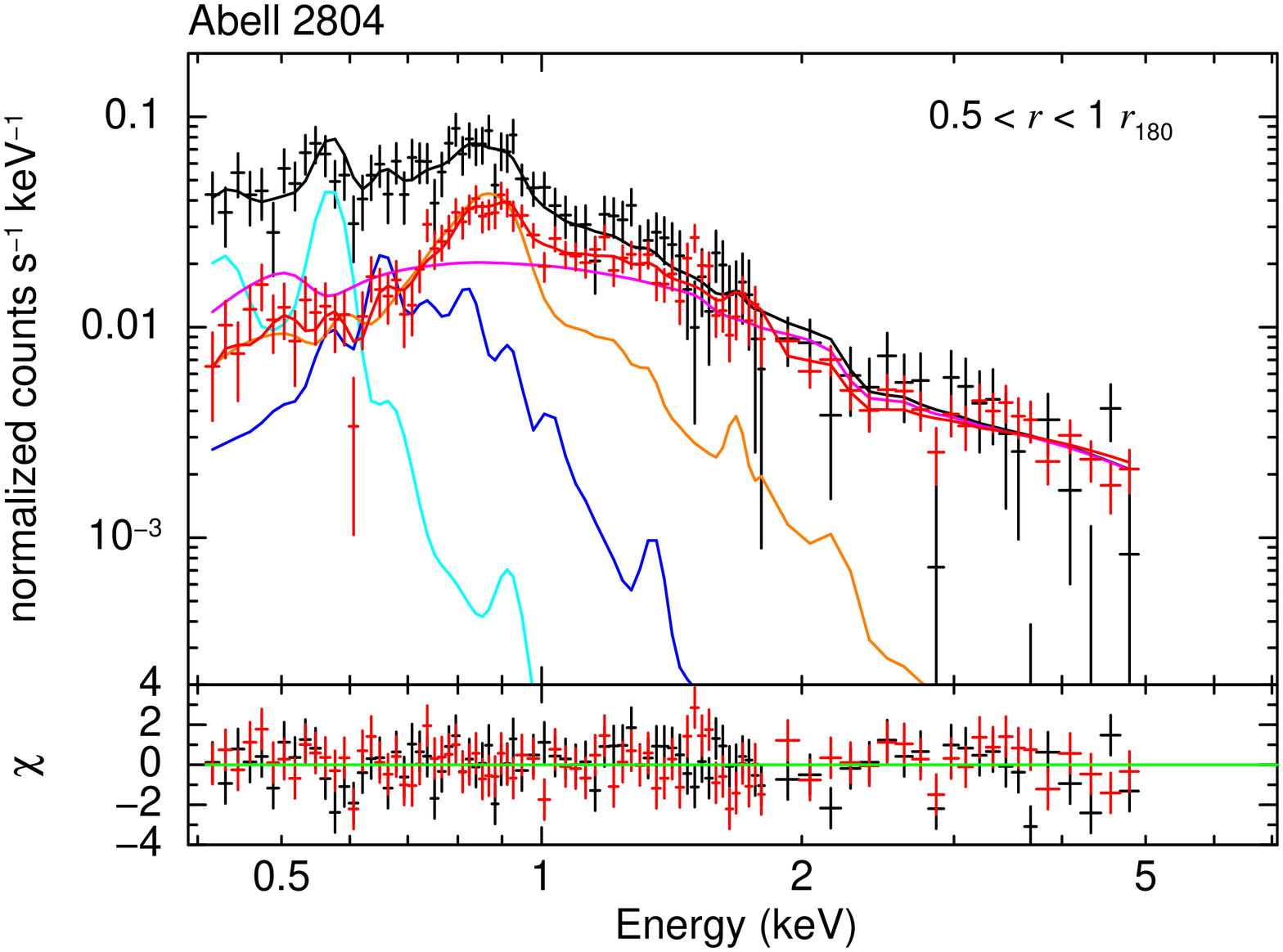}
\end{minipage}
\begin{minipage}{0.32\textwidth}
\FigureFile(\textwidth,\textwidth){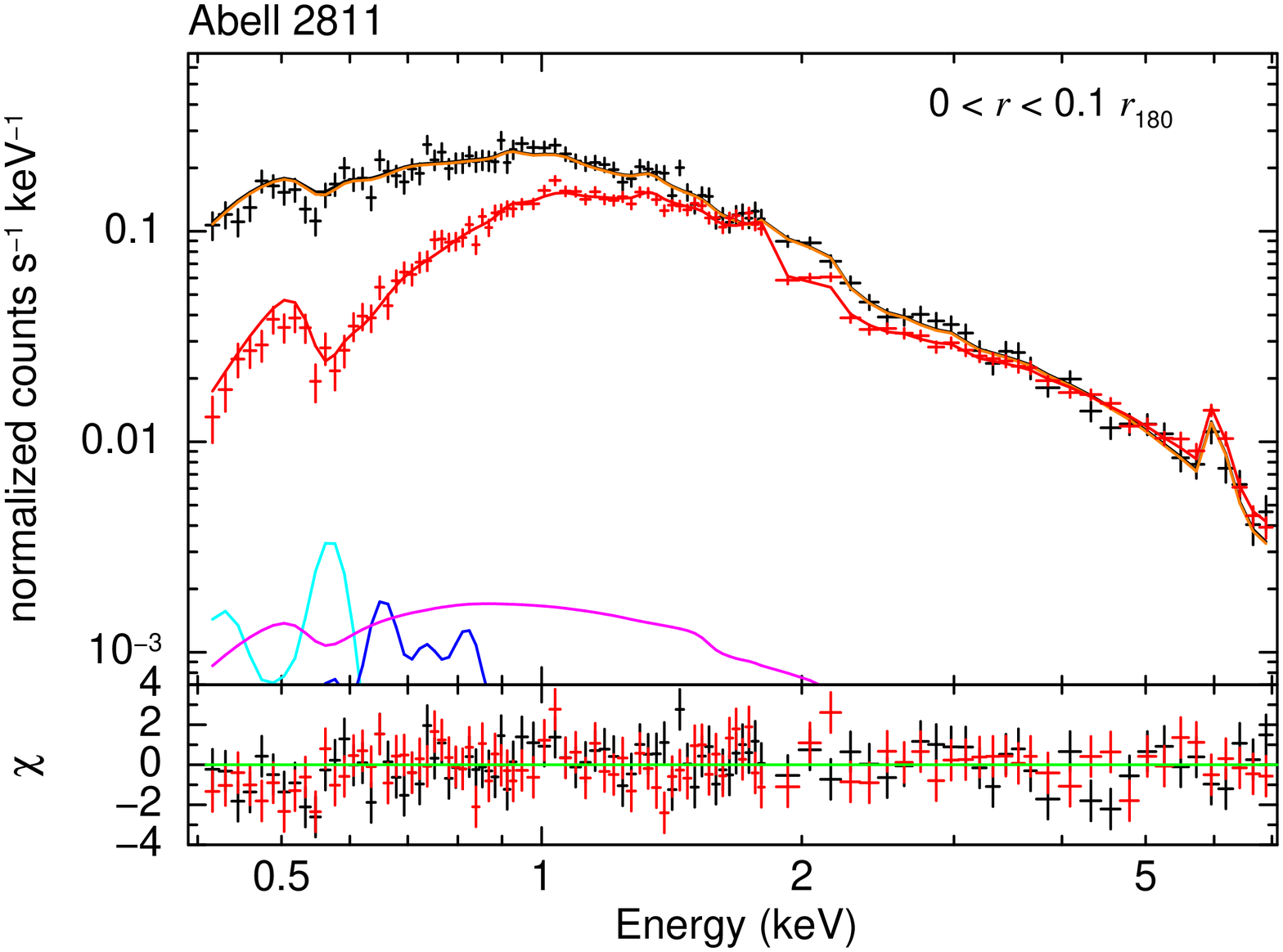}
\end{minipage}

\begin{minipage}{0.32\textwidth}
\FigureFile(\textwidth,\textwidth){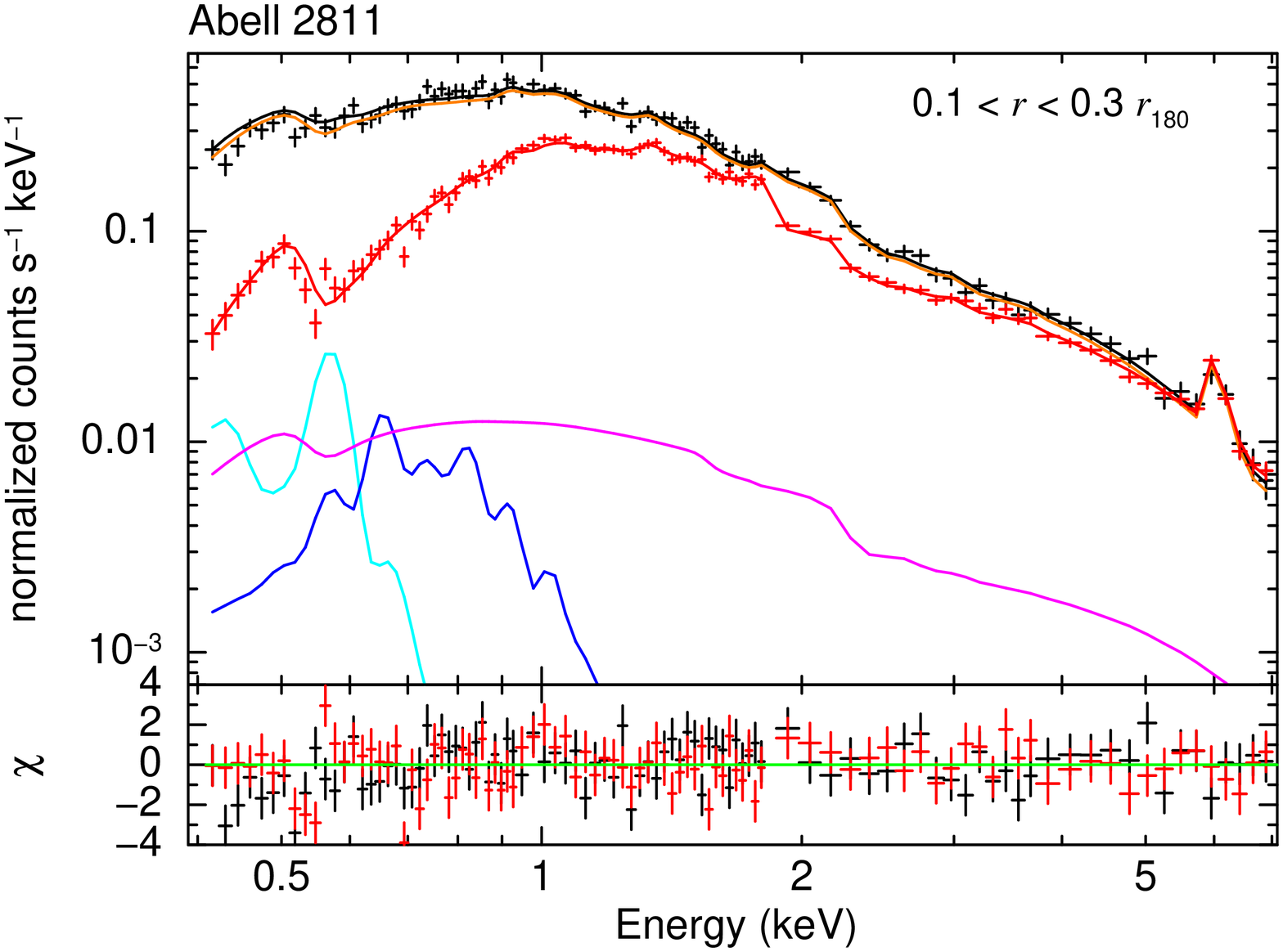}
\end{minipage}
\begin{minipage}{0.32\textwidth}
\FigureFile(\textwidth,\textwidth){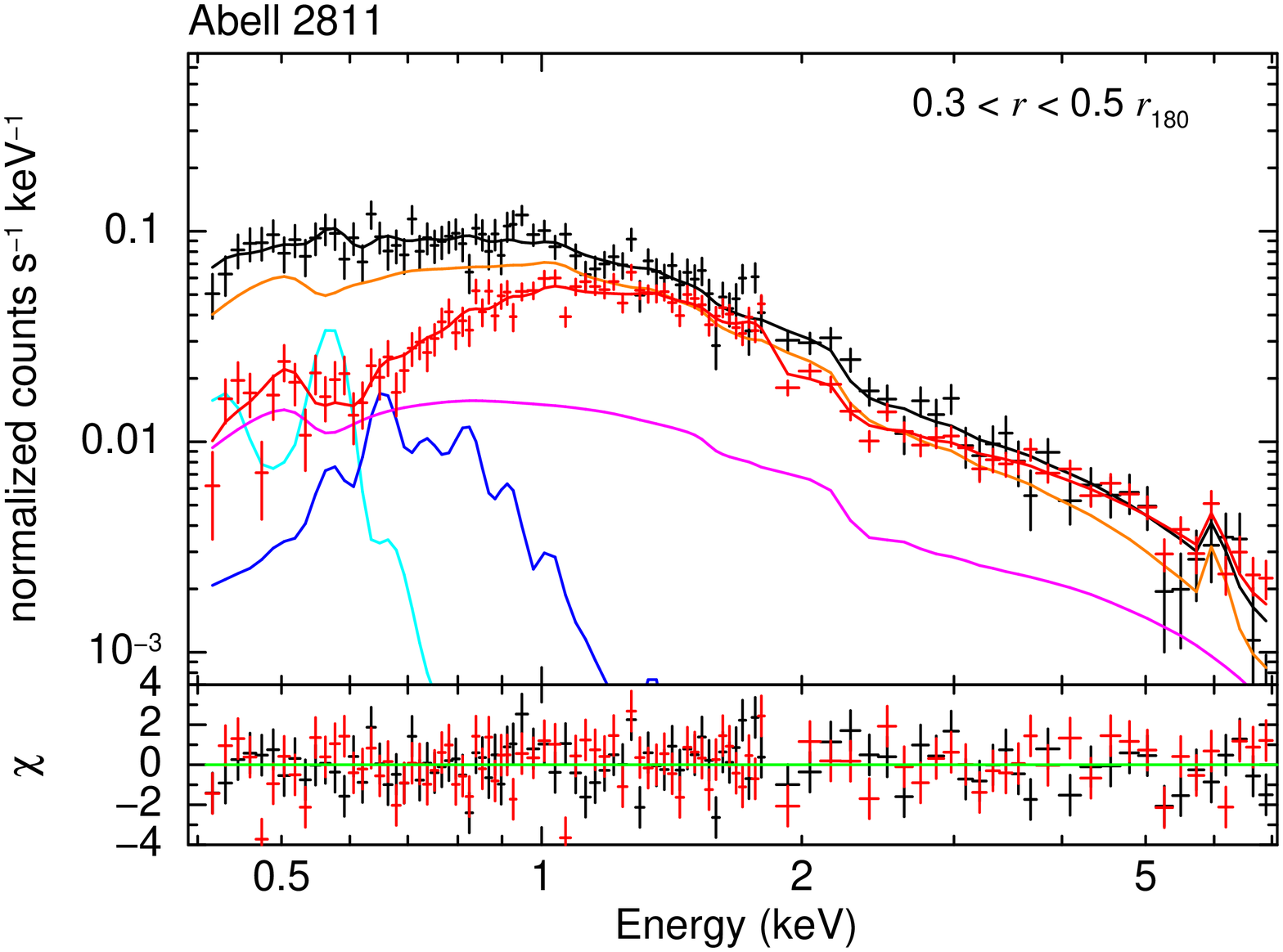}
\end{minipage}
\begin{minipage}{0.32\textwidth}
\FigureFile(\textwidth,\textwidth){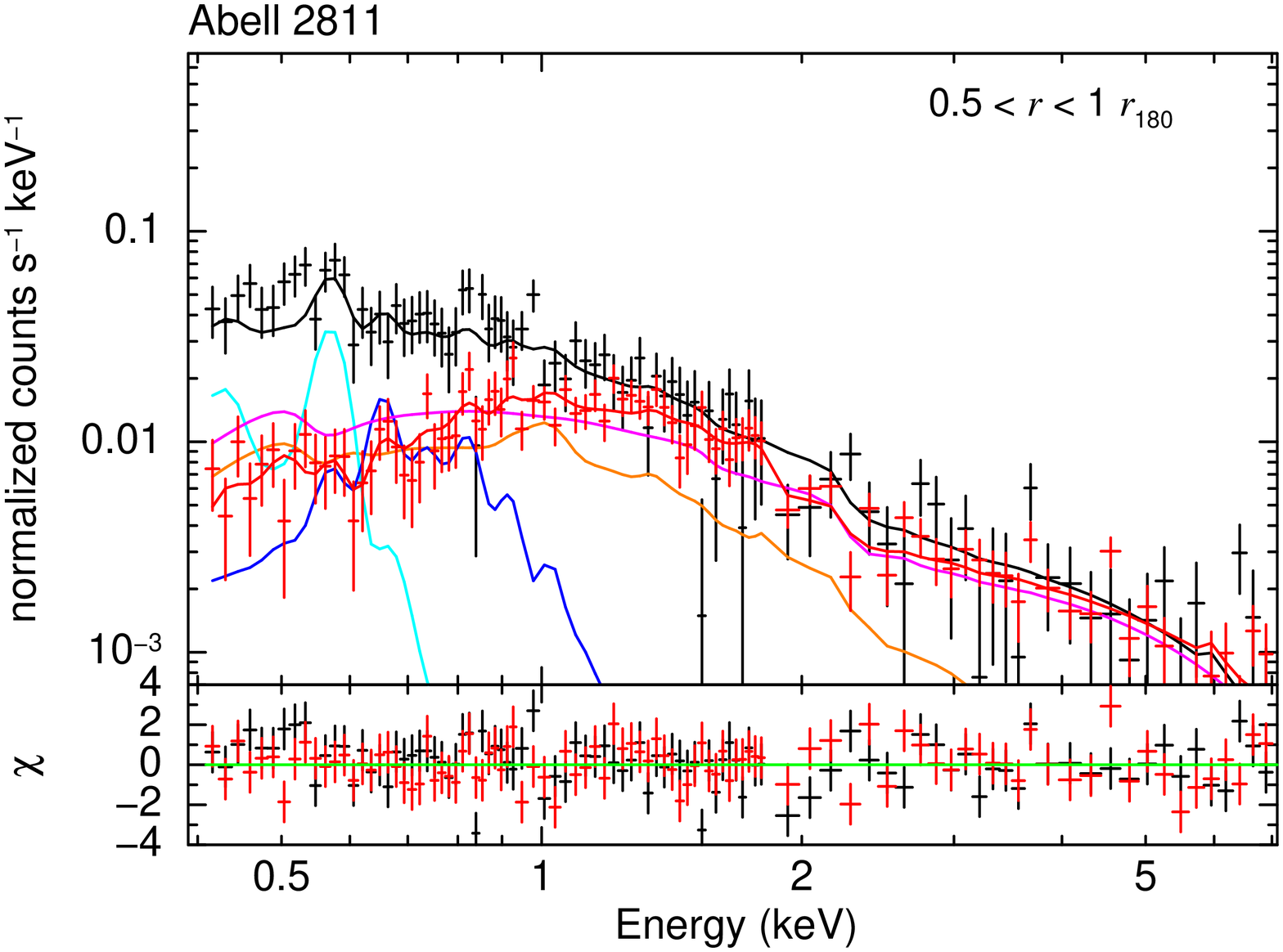}
\end{minipage}

\vspace*{-0.5ex}
\caption{
The panels show the observed spectra
for the regions of A2801, A2804, and A2811 which are denoted in the panels,
and the data are plotted by black and red crosses for BI and FI, respectively.
The estimated NXB components are subtracted,
and the black and red lines show the best-fit model 
for the BI and FI spectra, respectively.
The BI spectra of the ICM component are shown in an orange line. 
The CXB components are shown in a magenta line, 
and the Galactic emissions are indicated by cyan and blue lines, respectively.
For A2804, the additional ICM and LMXB components are shown in 
orange and magenta dashed lines, respectively.  
The energy range around the Si K-edge (1.825--1.840 keV) is ignored
in the spectral fits.
The lower panels show the fit residuals in units of $\sigma$.
}
\label{fig:5}
\end{figure*}

\begin{table*}
\caption{Summary of the parameters of the fits of A2801 and 
A2804.
Errors are 90\% confidence range of statistical errors, and do not 
include systematic errors. The solar abundance ratio of {\it angr} 
was assumed.}
\label{tab:7}
\begin{center}
\begin{tabular}{lccc} \hline
A2801 & & & \\
Region & 0$<r<0.4~r_{180}$ & 0.4$<r<0.6~r_{180}$ & 0.6$<r<1~r_{180}$ \\
$kT$ (keV) & 2.33$^{+0.15}_{-0.15}$ & 1.43$^{+0.24}_{-0.11}$ &
	     1.08$^{+0.44}_{-0.25}$ \\
$Z$ (solar) & 0.43$^{+0.12}_{-0.10}$ & 0.3 (fix) & 0.3 (fix)\\
$Norm^{\ast}$ & 15.9$^{+1.2}_{-1.1}$ & 7.01$^{+1.44}_{-1.30}\times10^{-1}$ & 6.51$^{+5.10}_{-4.32}\times10^{-2}$ \\
$\chi^2 / dof$ & 211/188 & 219/189 & 244/189 \\
\hline
\hline
A2804 & & & \\
Region & 0$<r<0.5~r_{180}$ & 0.5$<r<1~r_{180}$ & \\
$kT$ (keV) & 1.15$^{+0.31}_{-0.13}$/0.63$^{+0.23}_{-0.28}$ & 0.91$^{+0.06}_{-0.07}$ & \\
$Z$ (solar) & 0.3 (fix)& 0.3 (fix)& \\
$Norm^{\dagger}$ (10$^{-3}$) & 5.37$^{+1.02}_{-1.87}$/1.35$^{+2.19}_{-0.79}$ & 2.15$^{+0.27}_{-0.26}$ & \\
$\chi^2 / dof$ & 176/168 & 169/171 & \\
\hline
\\[-1ex]
\multicolumn{4}{l}{\parbox{0.7\textwidth}{\footnotesize
\footnotemark[$\ast$] 
Normalization of the {\it vapec} component scaled with a factor of
the selected region comparing to the assumed image in
 ``xissimarfgen'',
${\it Norm}= {factor} \int
n_{\rm e} n_{\rm H} dV \,/\, [4\pi\, (1+z)^2 D_{\rm A}^{\,2}]$ $\times
10^{-20}$~cm$^{-5}$~arcmin$^{-2}$, where $D_{\rm A}$ is the angular
distance to the source. }}\\
\multicolumn{4}{l}{\parbox{0.7\textwidth}{\footnotesize 
\footnotemark[$\dagger$] 
Normalization of the {\it apec} component
divided by the solid angle, $\Omega^{\makebox{\tiny\sc u}}$,
assumed in the uniform-sky ARF calculation (20$'$ radius),
${\it Norm} = \int n_{\rm e} n_{\rm H} dV \,/\,
(4\pi\, (1+z)^2 D_{\rm A}^{\,2}) \,/\, \Omega^{\makebox{\tiny\sc u}}$
$\times 10^{-14}$ cm$^{-5}$~400$\pi$~arcmin$^{-2}$, 
where $D_{\rm A}$ is the angular distance to the source.}}\\
\end{tabular}
\end{center}
\end{table*}
\begin{table*}
\caption{Summary of the parameters of the fits of A2811.
Errors are 90\% confidence range of statistical errors, and do not 
include systematic errors. The solar abundance ratio of {\it angr} 
was assumed.}
\label{tab:8}
\begin{center}
\begin{tabular}{rcccc} \hline
A2811 & & & & \\
Region & 0$<r<0.1~r_{180}$ & 0.1$<r<0.3~r_{180}$ & 0.3$<r<0.5~r_{180}$ 
& 0.5$<r<1~r_{180}$\\
$kT$ (keV) & 5.38$^{+0.22}_{-0.16}$ & 5.17$^{+0.12}_{-0.12}$ &
	     4.97$^{+0.39}_{-0.37}$ & 2.63$^{+0.82}_{-0.57}$ \\
O (solar) & 0.70$^{+0.60}_{-0.56}$ & 0.33$^{+0.40}_{-0.33}$ &
	     0.28$^{+0.13}_{-0.13}$ & 0.3 (fix) \\
Ne (solar) & 0.99$^{+0.66}_{-0.62}$  & 1.08$^{+0.47}_{-0.45}$ &
	     $\downarrow$ & $\downarrow$ \\
Mg, Al (solar) & 0.90$^{+0.67}_{-0.64}$ & 0.80$^{+0.48}_{-0.46}$ &
	     $\downarrow$ & $\downarrow$ \\
Si (solar) & 0.85$^{+0.47}_{-0.46}$ & 0.60$^{+0.34}_{-0.33}$ &
	     $\downarrow$ & $\downarrow$\\
S, Ar, Ca (solar) & 0.46$^{+0.46}_{-0.46}$ & 0.53$^{+0.33}_{-0.33}$ &
	     $\downarrow$ & $\downarrow$ \\
Fe, Ni (solar) & 0.34$^{+0.05}_{-0.05}$ & 0.35$^{+0.04}_{-0.04}$ &
	     $\downarrow$ & $\downarrow$ \\
$Norm^{\ast}$ & 648$^{+30}_{-30}$ & 81.5$^{+2.7}_{-2.7}$ &
	     8.08$^{+0.38}_{-0.38}$  & 0.968$^{+0.169}_{-0.167}$ \\
$\chi^2 / dof$ & 194/183 & 235/183 & 240/188 & 215/189\\
\hline
\\[-1ex]
\multicolumn{4}{l}{\parbox{0.7\textwidth}{\footnotesize
\footnotemark[$\ast$] 
Same as the unit of the normalization of A2801 in figure~\ref{fig:6}}}\\
\end{tabular}
\end{center}
\end{table*}
We also examined whether or not an additional apec component (with
redshift either zero or supercluster value) gave a significant
improvement in $\chi^2$ against a change of degree of freedom $\Delta
\nu$.  In the case of zero-redshift apec model, the improvement was
$\Delta \chi^2 = 1$ with $\Delta \nu = 3$ over the value of case~4 in
table \ref{tab:3}. This is not a significant change considering the
$F$-test.  On the other hand, in the case of adding a cluster-redshift
apec model ($z=0.108$), there was no improvement ($\Delta \chi^2 < 1$)
over the case~4 value, for a temperature range between 0.1--0.8 keV\@.

Since the fit is not formally acceptable
($\chi^2/\nu=231/170$ in case~4), we investigated the $\chi^2$
behavior in 2 energy bands, 0.4--1.0 and 1.0--5.0 keV, separately.
As a result, the $\chi^2$-value in 1.0--5.0 keV (the reduced $\chi^2
\sim 1.5$) was much larger than that in 0.4--1.0 keV (the reduced
$\chi^2 \sim 1.0$). Changing the parameters of the CXB component did
not improve the fit.  The spectral data in this high-energy range
seem to fluctuate from bin to bin by more than the statistical
limit. There are some features around the edges of Si-K (1.8 keV)
and Au-L (3.4 keV), suggesting a problem in the XIS response
function. However, the model reproduces average feature of the
observed spectrum quite well, yielding the CXB spectrum consistent
with the previous \citet{kushino02} result, and the fit is good in
the energy range of our interest ($< 1$ keV). Therefore, we proceeded
with the spectral fit using these data.

We also investigated the effect of possible incorrect calibration,
i.e., NXB level and OBF contaminations, by artificially changing the
values by $\pm10$\% and comparing the resultant $\chi^2$ value.  While
the temperatures of the LHB and MWH did not change within $<1$\%
compared with the values for the case~1 in table~\ref{tab:3}, the
normalization and $\Gamma$ of the CXB models changed by
$\sim$10\%\@. We also examined the other uncertainty in the OBF
contaminant by changing the absorber thickness by $\pm10\%$.  The
temperatures of LHB and MWH and $\Gamma$ and normalization of the CXB
did not change within $<1\%$, the normalizations of the LHB and MWH
changed by $\pm \sim10$\%\@.  A list of $\chi^2$/dof by changing the
systematic errors in case 4 is presented in table~\ref{tab:4}\@.

We also estimated the fraction of photons entering from outside of the
extracted regions using a simulator of the Suzaku XRT/XIS system
``xissim''.  The photon contamination can cause a significant effect
in the case of cluster observations, since much bright center is
outside of the field and the point-spread function of the Suzaku XRT
shows an extended tail (see e.g.\ \cite{sato07a}).  If we assume a
beta-model of A2811 emission as described in subsection
\ref{subsec:data} extending beyond the virial radius, a half of the
simulated photons in the extracted region, which is outside of the
virial radius, is originated from outside of the extracted
region. However, this emission has a flux of $\sim1/50$ of the
estimated CXB level below 2 keV\@.  Thus, the effect of the stray
light is negligible in the spectral fits.

\subsection{Observations close to the Sculptor supercluster}

In order to estimate the Galactic and CXB spectra precisely, we looked
into other observations which were carried out near the Sculptor
supercluster.  There are two observations: NGC~253 offset region, and
Q0109-3518\@. Their observation logs are shown in table \ref{tab:5}.
We analyzed these data with the same criteria used for the fore- and
background spectral estimation as described in subsection
\ref{subsec:bgd}.  As mentioned in subsection \ref{subsec:data}, the
contamination on the XIS optical blocking filter has increased after
the launch, and the effective area in the energy range below 1 keV
were smaller in these offset data as shown in figure~\ref{fig:4}.
However, because this effect was properly included in the ARF, the
observations were useful in improving the background estimation.

After subtracting point-like sources as shown in figure \ref{fig:3},
we carried out spectral fits with the same model, consisting of LHB,
MWH and CXB, by changing the parameters to be either free or fixed.
The resultant parameters are shown in table \ref{tab:6} and in figure
\ref{fig:4}.  As a result, the spectra of these near-by regions are
also well-fitted with the background model as shown in
subsection~\ref{subsec:bgd}. The intensities of the Galactic and CXB
components are consistent with those for the region outside of the
virial radius of the clusters in the Sculptor supercluster.

\subsection{Radial Profile of ICM Properties for each cluster}
\label{subsec:spec}
\begin{figure}[ht]
\begin{center}
\FigureFile(0.4\textwidth,0.4\textwidth){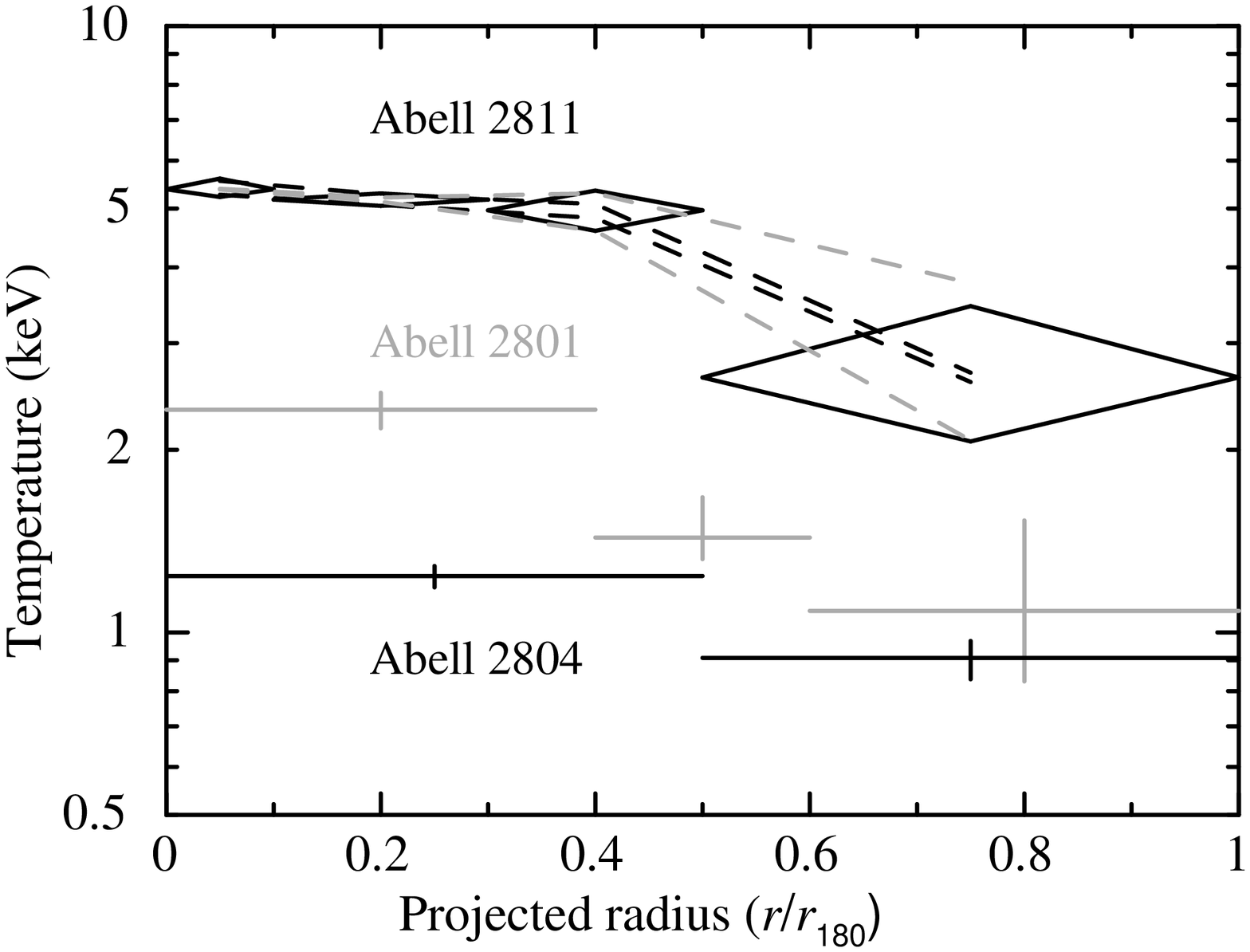}
\caption{
Radial temperature profiles of A2811, A2801, and A2804 
against the projected radius in unit of the virial radius. 
Black and Light gray lines show typical systematic changes of the 
best-fit values by varying the thickness of the OBF contaminant and 
NXB levels by $\pm10$\%, respectively. }
\label{fig:6}
\end{center}
\end{figure}

We extracted spectra from the regions $0<r<0.1~r_{180},
0.1<r<0.3~r_{180}, 0.3<r<0.5~r_{180},$ and $0.5<r<1~r_{180}$, centered
on (RA, Dec) = (\timeform{0h42m09.3s}, \timeform{-28D32'33''}) for
A2811, $0<r<0.4~r_{180}, 0.4<r<0.6~r_{180},$ and
$0.6<r<1~r_{180}$, centered on (RA, Dec) = (\timeform{0h38m32.3s},
\timeform{-29D04'44''}) for A2801, and 0$<r<0.5~r_{180}$,
$0.5<r<1.0~r_{180}$ centered on (RA, Dec) = (\timeform{0h39m38.7s},
\timeform{-28D53'21''}) for A2804, respectively.  The spectra
with BI and FI for all regions were fitted simultaneously.  Energy
ranges used for the fit are 0.4--7.1 keV for A2811 and
A2801, and 0.4--5.0 keV for A2804, respectively.  In the
simultaneous fit, the Galactic and CXB components based on the case 4
values as described in subsection~\ref{subsec:bgd} were commonly
included for all regions.  We excluded the narrow energy band around
Si K-edge (1.825--1.840~keV) because its response was not modeled
correctly.  The energy range below 0.4~keV was also excluded because
the C edge (0.284~keV) seen in the BI spectra could not be reproduced
well in our data. These are the known problems in the XIS response.

The BI and FI spectra of the ICM in A2801 and A2811 were
well-represented by a single temperature model as shown in
figure~\ref{fig:5}.  The spectra except for the $r<0.3~r_{180}$ region
of A2811 were all fitted with {\it apec} model, while these inner
region spectra of A2811 needed fitting with {\it vapec} model.
In this fit, the metal abundances were linked in the following way,
Mg=Al, S=Ar=Ca, and Fe=Ni.  As for A2804, the spectrum in
$0.5<r<1~r_{180}$ was represented by a single temperature model, but
the data in the inner $r<0.5~r_{180}$ region needed an additional low
mass X-ray binary (LMXB) component represented by {\it zbremss} with
$kT = 10$ keV as well as the {\it apec} ICM component. The A2804
emission is significantly affected by point-like sources due to their
lower ICM temperature than those in other clusters.  Results of the
temperature and abundance from the spectral fit for individual regions
are summarized in table~\ref{tab:7}--\ref{tab:8} and
figure~\ref{fig:6}. The typical systematic error on the temperature of
A2811 due to the OBF contamination and change of NXB level by
$\pm10\%$ are indicated.

The ICM temperatures of the clusters clearly decrease towards the
virial radii. The temperatures in 0.5--1~$r_{180}$ region drop to
about 60\% of the peak temperature in the three clusters.  This
tendency is consistent with the previous Suzaku results
\citep{bautz09,reiprich09,hoshino10}\@.  While the metal abundances of
$\alpha$ elements in the central region of A2811 are about the solar
value, the iron abundance is $\sim 0.4$ times the solar value (see
table~\ref{tab:8}).  The obtained temperatures are consistent with the
ASCA results \citep{obayashi98}, but the present results give better
accuracy and radial distribution of temperature. The metal abundance
in the central region of A2801 is also 0.4 solar, which is mostly
determined by iron.  The metal abundance in the outer region of A2801
and the whole region of A2804 had to be fixed at 0.3 solar because of
the poor photon statistics.  The iron abundances of these clusters in
the $r<0.1~r_{180}$ region may be slightly lower than those in other
clusters (e.g.\ \cite{tamura08,hoshino10}), while the alpha elements
in A2811 show some enhanced abundance compared with the iron.

\citet{kawaharada10} reported a higher temperature and entropy region
in the northeastern edge of A1689 which was connected to an overdense
filamentary structure of galaxies outside of the cluster. As shown in
figure~\ref{fig:1} right, the galaxy distribution around A2811 shows
an extension to the northwest direction rather than the southwest
direction. We investigated the difference of temperature between these
directions in the radius range of $0.3<r<0.5~r_{180}$ and
$0.5<r<1~r_{180}$ from A2811\@. Metal abundance was fixed to 0.3 solar
because of the poor statistics.  As a result, we found no temperature
variation between the two directions within the statistical error.

\subsection{Upper limit of Oxygen emission lines}
\label{subsec:limit}

\begin{figure}
\begin{center}
\FigureFile(0.4\textwidth,0.4\textwidth){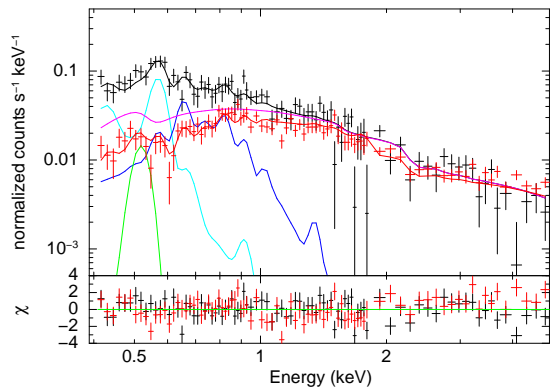}
\caption{
The panels show the resultant fits for the constraint of the 
intensities of the O\emissiontype{VII} and O\emissiontype{VIII} 
lines in the filament region between clusters of the Sculptor supercluster.
The O\emissiontype{VII} emission line is shown in a green line, 
and the notations of the other lines are same as shown in figure~\ref{fig:2}. 
}
\label{fig:7}
\end{center}
\end{figure}

Finally, we fitted the spectra taken outside of the virial radii of
the clusters. The region is the A2811 offset region, supposed to be
the filament connecting the clusters. This region is the same as the
one we used to estimate the fore- and background emissions as shown in
subsection \ref{subsec:bgd}.  The spectrum in this
region is well-represented by the Galactic and CXB components
without an additional model required as shown in
subsection~\ref{subsec:bgd}, however we investigate the upper limit
of the intensities of the additional O\emissiontype{VII} and
O\emissiontype{VIII} lines at the cluster redshift from the WHIM
origin.  We fitted the data with the following model: apec$_{\rm
  LHB}+{\rm phabs}\times({\rm apec}_{\rm MWH}+{\rm pow}_{\rm CXB} +
{\rm gaussian}_{\rm O\emissiontype{VII}} + {\rm gaussian}_{\rm
  O\emissiontype{VIII}})$\@.  Here, the parameters of the Galactic and
CXB components followed our previous fit in case~5. The temperatures
of the Galactic components and the photon index of the CXB component
were fixed, and the normalizations were free in the fit. The
redshifted central energies of the O\emissiontype{VII} and
O\emissiontype{VIII} gaussian lines were fixed at 518 and 589 eV\@,
respectively, with no intrinsic width of the lines assumed.
We employed the increment of $\Delta \chi^2 =4$ as the
measure for the $2\sigma$ upper limits of the line intensities. This
value also corresponds to the 95\% limit for an additional single
parameter (single line intensity) in the $F$ test. As a result, we
determined the upper limit of the O\emissiontype{VII} and
O\emissiontype{VIII} lines to be 8.1 and 5.1 $\times10^{-8}$ photons
cm$^{-2}$ s$^{-1}$ arcmin$^{-2}$ as the $2\sigma$ confidence limits as shown in
table \ref{tab:9} and figure \ref{fig:7}.  Even if we allowed the
redshift of the O\emissiontype{VII} and O\emissiontype{VIII} lines to
vary to give the peak energies in 490--540 and 560--620 eV,
respectively, the upper limit did not change within the statistical
errors.  Our resultant values are almost consistent or give slightly
stronger constraint compared with the results in \citet{takei07b}.
This is due to the combination of less coupling between the redshifted
and the Galactic O lines, to slightly lower Galactic foreground
intensity and to lower interstellar absorption.

\begin{table}
\caption{Resultant intensities of O\emissiontype{VII} and
 O\emissiontype{VIII} emission lines with 2$\sigma$ confidence range.}
\label{tab:9}
\begin{center}
\begin{tabular}{lc} \hline 
O\emissiontype{VII} & \\
Center (keV) & 0.518(fixed) \\
Sigma & 0 (fixed) \\
I ($\times10^{-8}$ photons cm$^{-2}$ s$^{-1}$ arcmin$^{-2}$) & 4.2$^{+3.9}_{-3.9}$ \\
\hline
O\emissiontype{VIII} & \\
Center (keV) & 0.589(fixed) \\
Sigma & 0 (fixed) \\
I ($\times10^{-8}$ photons cm$^{-2}$ s$^{-1}$ arcmin$^{-2}$) & 0.0$^{+5.1}_{-0.0}$ \\
\hline\\
\end{tabular}
\end{center}
\end{table}

\begin{figure}[th]
\begin{center}
\FigureFile(0.45\textwidth,0.45\textwidth){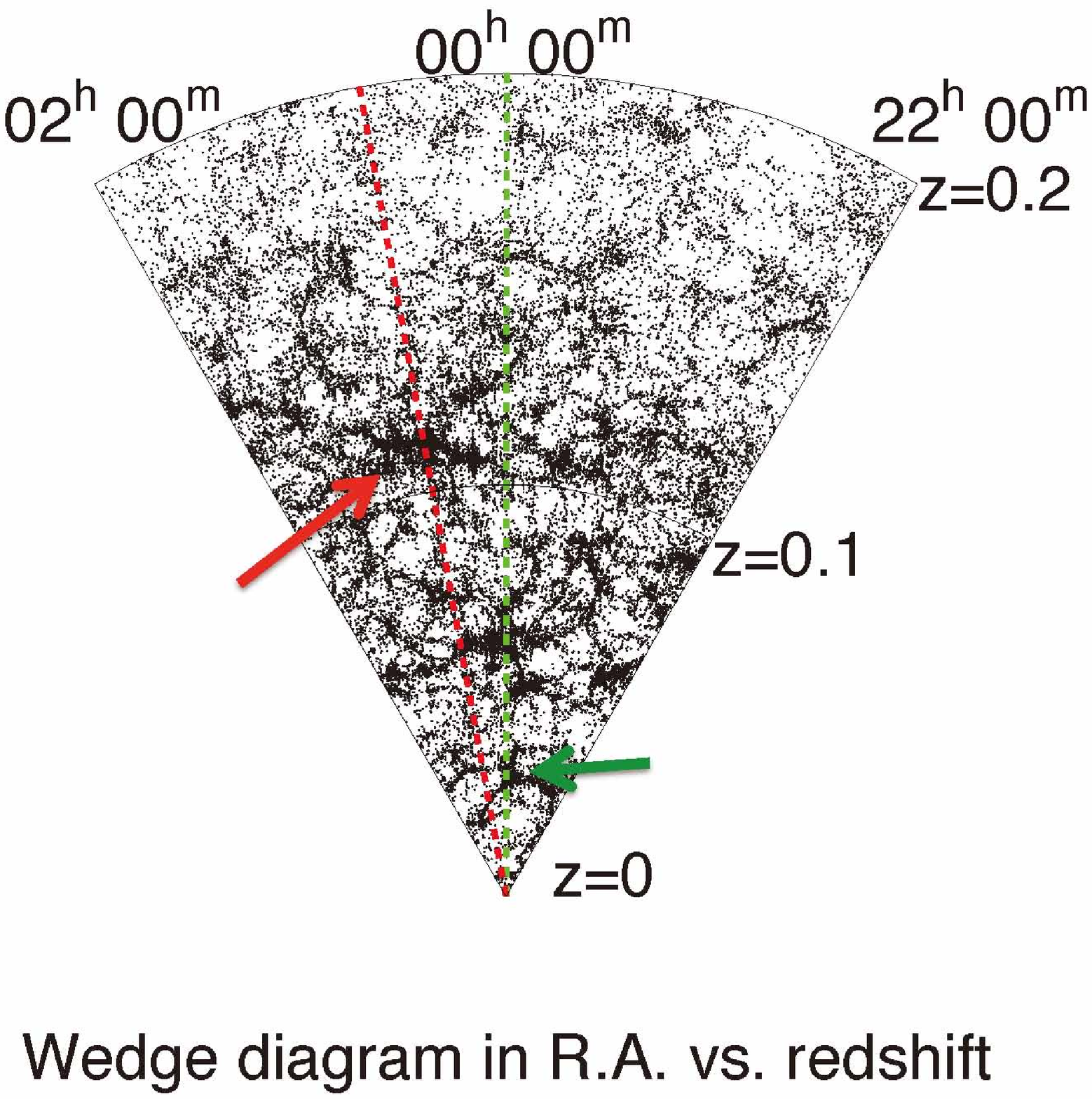}
\caption{
Wedge diagram expressed in R.A. vs. redshift projected 
on $-32^{\circ} < {\rm Dec.} < -26^{\circ}$.
The data are taken from 2dFGRS as mentioned in subsection
 \ref{subsec:obs} and figure \ref{fig:1}.
About 44,000 galaxies exist in the direction of the map and 
the redshift, $z<0.2$\@. The red and green dashed line
show the direction of our observations
and ``the Sculptor Wall'' at $z\sim0.03$ in 
\citet{fang10}, respectively.
The red and green arrows also show the Sculptor supercluster and 
``the Sculptor Wall'', respectively.
}\label{fig:8}
\end{center}
\end{figure}

In order to constrain the WHIM density based on the O line upper
limits, we show a wedge diagram of galaxy distribution in figure
\ref{fig:8}. This is expressed in R.A. vs.\ redshift projected for the
declination range $-32^{\circ}<{\rm Dec.}<-26^{\circ}$ in the
direction of the Sculptor supercluster. The galaxy concentration
around the Sculptor supercluster around $z\sim0.11$ is obvious and we
assume $\sim 25$ Mpc for the line-of-sight depth of the WHIM
distribution, which corresponds to $0.105<z<0.111$.  As mentioned in
\citet{takei07b}, we constrain the density of the WHIM cloud assuming
the temperature to be $T = 2\times10^6$ K\@. The line intensity,
$I<8.1\times 10^{-8}$ photons cm$^{-2}$ s$^{-1}$ arcmin$^{-2}$ at
z=0.1079, and electron-to-hydrogen number density ratio $n_e/n_{\rm
  H}=1.2$ for ionized gas give the following condition:
\begin{eqnarray}
n_{\rm H} &<& 1.6\times 10^{-5}~{\rm cm}^{-3}~(Z/0.1~Z_{\odot})^{-1/2}~
(L/25~{\rm Mpc})^{-1/2}.
\label{eq:1}\end{eqnarray}
The corresponding overdensity $\delta \equiv n_{\rm H}/\bar{n}_{\rm
  H}$ is
\begin{eqnarray}
\delta&<&60~ (Z/0.1~Z_{\odot})^{-1/2}~
(L/25~{\rm Mpc})^{-1/2}.
\label{eq:2}\end{eqnarray}

Compared with the previous Suzaku observations \citep{takei07b,
  tamura08}, our upper limits for the O\emissiontype{VII} and
O\emissiontype{VIII} lines imposed slightly tighter constraints on the
WHIM emission.  The present result is consistent with
the expected values from numerical simulations, indicating a baryon
over-density $\rho / \rho_{\rm crit} \sim 10-100$ by \citet{dolag06}
for example.  We note that the dense part of WHIM is found only in
the superclusters and the gas is likely to be inhomogeneous with a
wide range of temperature and density. In the case of the bridge
region between A222 and A223, Werner et al.\ (2008) showed the
temperature to be ~0.9 keV\@. Such a high temperature (also with
high density) yields lower emissivity of O\emissiontype{VII} and
O\emissiontype{VIII} lines.  Thus, estimation of temperature and
density is not straightforward.

As mentioned in section \ref{sec:intro}, \citet{fang10} reported a
significant detection of O\emissiontype{VII} absorption line in the
direction of the ``Sculptor Wall'' at $z\sim 0.03$.
\citet{zappacosta10} also searched for the C, O, and Ne absorption
lines from the ``Pisces--Cetus Supercluster'' and the `` Farther
Sculptor Wall'' in the same line of sight as the ``Sculptor Wall''.
Even though the Sculptor supercluster and the above studied filaments
are not physically connected with each other, we can compare the
estimated hydrogen density.  Our upper limit on the density is twice
larger than the ``Sculptor Wall'' value reported by \citet{fang10}.
Also, the overdensities derived for the ``Pisces--Cetus Supercluster''
and ``Farther Sculptor Wall'' are $\sim 1/4$ times of our result.
This indicates a higher sensitivity of the absorption line measurement
in the detection of WHIM\@. In fact, Suzaku cannot probably detect the
O line emission from the ``Sculptor wall'' because the redshift is too
low to separate from the Galactic emission lines.  Thus, future
quantitative studies of the WHIM will need X-ray optics with larger
effective area and instruments with much higher energy resolution such
as microcalorimeters, which would play a key role in future
investigation of the cosmic plasma diagnostics.

\section{Summary}

Suzaku carried out observations of the Sculptor supercluster in four
fields: A2811, A2811 offset, A2804, and A2801 regions.  After
estimation of the foreground and background emission, we examined the
annular cluster spectra to the virial radius.  The resultant
temperature of the clusters showed significant decrease from the
central region toward the virial radius, consistent with the previous
report for other clusters.  Iron abundance suggests a nearly uniform
distribution.  We also searched for the oxygen line emission outside
of the virial radii of the clusters, corresponding to the filament
region in the Sculptor supercluster.  The spectral fits on the
filament region data did not require an additional component except
for the Galactic and CXB components. We constrained the intensities of
redshifted O\emissiontype{VII} and O\emissiontype{VIII} lines to be
less than 8.1/5.1 $\times 10^{-8}$ photons cm$^{-2}$ s$^{-1}$ arcmin$^{-2}$,
respectively, as the $2\sigma$ upper limits.  The intensity of
O\emissiontype{VII} line indicates $n_{\rm H}< 1.6\times 10^{-5}$
cm$^{-3}$ ($Z/0.1~Z_{\odot}$)$^{-1/2}$ ($L/25~{\rm Mpc}$)$^{-1/2}$,
and the over density, $\delta<60$ ($Z/0.1~Z_{\odot}$)$^{-1/2}$
($L/25~{\rm Mpc}$)$^{-1/2}$\@.

\bigskip
Authors are grateful to Noel Bermann for supporting the data analysis.
We also thank the referee for providing valuable comments.
K.S is supported by a JSPS Postdoctral fellowship for research abroad.
Part of this work was financially supported
by the Ministry of Education, Culture, Sports, Science
and Technology, Grant-in-Aid for Scientific Research Nos.
20340068, 20340041, 20840051, 21224003, 21740134.

\end{document}